\documentclass[12pt]{article}
\usepackage{url}
\usepackage{amssymb,amsmath,bm}%
\usepackage{amsfonts}%
\usepackage{geometry}
\usepackage{setspace}
\newcommand{\beaa}{\begin{eqnarray*}}
\newcommand{\eeaa}{\end{eqnarray*}}
\newcommand{\bea}{\begin{eqnarray}}
\newcommand{\eea}{\end{eqnarray}}

\usepackage{amsmath}
\usepackage{hyperref}
\usepackage{array} 
\usepackage{bm}
\newcommand\independent{\protect\mathpalette{\protect\independenT}{\perp}}
\def\independenT#1#2{\mathrel{\rlap{$#1#2$}\mkern2mu{#1#2}}} 
\usepackage{tabularx}
\usepackage{multirow}
\usepackage{dcolumn}
\numberwithin{equation}{section}
\renewcommand{\theequation}{\thesection.\arabic{equation}}

\newtheorem{thm}{Theorem}[section]
\renewcommand{\thethm}{\arabic{section}.\arabic{thm}}
\newtheorem{cor}[thm]{Corollary}
\newtheorem{lemma}[thm]{Lemma}

\newtheorem{rmk}[thm]{Remark}
\newtheorem{ass}[thm]{Assumption}

\usepackage{graphicx}
\usepackage{float}
\usepackage{rotating}
\begin{document}

\title{\bf{Central Quantile Subspace}}
\date{}
\maketitle

\begin{center}
{\Large{\textbf{Eliana Christou} \footnote{\label{note1}This work was supported, in part, by funds provided by the University of North Carolina at Charlotte.}\\
\textbf{University of North Carolina at Charlotte}}}
\end{center}

\noindent \textbf{Abstract:} Quantile regression (QR) is becoming increasingly popular due to its relevance in many scientific investigations.  There is a great amount of work about linear and nonlinear QR models.  Specifically, nonparametric estimation of the conditional quantiles received particular attention, due to its model flexibility.  However, nonparametric QR techniques are limited in the number of covariates.  Dimension reduction offers a solution to this problem by considering low-dimensional smoothing without specifying any parametric or nonparametric regression relation.  Existing dimension reduction techniques focus on the entire conditional distribution.  We, on the other hand, turn our attention to dimension reduction techniques for conditional quantiles and introduce a \textit{new method} for reducing the dimension of the predictor $\mathbf{X}$.  The \textit{novelty} of this paper is threefold.  We start by considering a \textit{single index quantile regression} model, which assumes that the conditional quantile depends on $\mathbf{X}$ through a \textit{single} linear combination of the predictors, then extend to a \textit{multi index quantile regression} model, and finally, generalize the proposed methodology to any \textit{statistical functional} of the conditional distribution.  The performance of the methodology is demonstrated through simulation examples and a real data application.  Our results suggest that this method has a good finite sample performance and often outperforms existing methods.      


\section{Introduction} \label{Intro}

Quantile regression (QR) was first introduced by Koenker and Bassett (1978) and since then, it has received a lot of attention.  There is a considerably large body of work about QR models.  Specifically, Koenker and Bassett (1978) considered the linear QR model, while, among others, Chaudhuri (1991), Yu and Jones (1998), Kong et al. (2010), and Guerre and Sabbah (2012), considered the completely flexible nonparametric QR model.  However, when the number of the predictors is large, nonparametric methods require smoothing over a high-dimensional space, where the data becomes sparse.  Dimension reduction offers a solution to this problem by considering low-dimensional smoothing without specifying any parametric or nonparametric regression relation.

Existing dimension reduction techniques focus on the entire conditional distribution and include, among others, sliced inverse regression (SIR, Li 1991), principal Hessian directions (pHd, Li 1992), sliced average variance estimation (SAVE, Cook and Weisberg 1991), parametric inverse regression (PIR, Bura and Cook 2001), minimum average variance estimation (MAVE, Xia et al. 2002), partial SIR (Chiaromonte et al. 2002), contour regression (Li et al. 2005), directional regression (DR, Li and Wang 2007), sliced regression (SR, Wang and Xia  2008), and more recently, sliced inverse median regression (SIMR, Christou 2018).  See also Hristache et al. (2001), Li and Dong (2009), Dong and Li (2010), Yin and Li (2011), Zhang et al. (2011), Ma and Zhu (2012), Shin and Artemiou (2017), and Zhu et al. (2017).

When specific aspects of the conditional distribution are of interest, such as the conditional mean, conditional variance, and conditional quantile of the response given the covariates, the above methods are inefficient as they focus on the entire conditional distribution and provide more directions than necessary.  Recent research focuses on these statistical functionals.  Specifically, Cook and Li (2002) focused on the conditional mean and introduced the central mean subspace (CMS), Yin and Cook (2002) focused on the conditional $k$th moment and introduced the central $k$th moment subspace (CKMS), and Zhu and Zhu (2009) focused on the conditional variance and introduced the central variance subspace (CVS).  

In this work we focus on dimension reduction techniques for \textit{conditional quantiles} and propose a new method for finding the fewest linear combinations of $\mathbf{X}$ that contain all the information on that function.  Existing literature considers a single index quantile regression (SIQR) model and proposes an iterative algorithm for estimating the vector of the coefficients of the linear combination of $\mathbf{X}$ (Wu et al. 2010, Kong and Xia 2012).  To avoid iterations and convergence issues, Christou and Akritas (2016) proposed a non-iterative algorithm for estimating that vector of coefficients.  While their methodology has been extended to allow for simultaneous variable selection and parameter estimation (Christou and Akritas 2018), it is still limited to a small number of covariates.  Kong and Xia (2014) proposed an adaptive composite QR approach, which can be used for estimating multiple linear combinations of $\mathbf{X}$ that contain all the information about the conditional quantile, while Luo et al. (2014) introduced a sufficient dimension reduction method that targets any statistical functional of interest, including the conditional quantile.  The authors proposed an efficient estimator for the semiparametric estimation problem of this type.  However, while their work covers a wide range of applications, the finite sample performance of the efficient estimator is not necessarily the best.  For the above reasons, a different approach is needed.  

In this paper, we introduce the \textit{$\tau$th central quantile subspace} ($\tau$-CQS; see Section \ref{Sec2}) and propose an algorithm for estimating it.  Specifically, in Section \ref{Sec3} we focus on a SIQR model and propose a method for estimating the one-dimensional $\tau$-CQS, in Section \ref{Sec4} we extend the methodology to a multi index quantile regression (MIQR) model, while in Section \ref{Sec5} we generalize the proposed methodology to any statistical functional of interest.  In Section \ref{Sec6} we present a brief discussion on the estimation of the dimension of the $\tau$-CQS, and in Section \ref{Sec7} we present results from several simulation examples and a real data application.  A discussion  is given in Section \ref{Sec8}.  The assumptions, some lemmas, and the proof of Theorems \ref{thm5.1} and \ref{thm5.2} are given in the Appendix.

\section{The $\tau$th Central Quantile Subpsace} \label{Sec2}

We start by recalling some basic definitions from Li (1991).  Let $Y$ and $\mathbf{X}$ denote a univariate response and a $p \times 1$ vector of predictors, respectively, and let $\mathbf{A}=(\boldsymbol{\alpha}_{1}, \dots, \boldsymbol{\alpha}_{d})$ denote a $p \times d$ matrix, where $\boldsymbol{\alpha}_{1}, \dots, \boldsymbol{\alpha}_{d}$ are column vectors and $d \leq p$.  Assume that $Y \independent \mathbf{X} | \mathbf{A}^{T} \mathbf{X}$, i.e., that $Y$ and $\mathbf{X}$ are independent given $\mathbf{A}^{T} \mathbf{X}$.  This means that, the $d \times 1$ predictor vector $\mathbf{A}^{T}\mathbf{X}$ captures all we need to know about $Y$, implying that we can replace the $p \times 1$ predictor vector $\mathbf{X}$ with the $d \times 1$ predictor vector $\mathbf{A}^{T}\mathbf{X}$ without loss of information.  The space spanned by the column vectors $\boldsymbol{\alpha}_{1}, \dots, \boldsymbol{\alpha}_{d}$, denoted by $\mathcal{S}(\mathbf{A})$, is called the \textit{dimension reduction subspace} for the regression of $Y$ on $\mathbf{X}$.  The greatest dimension reduction in the predictor vector is achieved using the smallest dimension reduction subspace, called the \textit{central subspace} (CS), and denoted by $\mathcal{S}_{Y|\mathbf{X}}$.

A straightforward extension of the CS to conditional quantiles, and a special case of Definition 1 of Luo et al. (2014), states the following.  Let, for $\tau \in (0,1)$, $Q_{\tau}(Y|\mathbf{x})\equiv Q_{\tau}(Y|\mathbf{X}=\mathbf{x})=\inf \{y: \Pr (Y \leq y | \mathbf{X}=\mathbf{x}) \geq \tau\}$ denote the $\tau$-th conditional quantile of $Y$ given $\mathbf{X}=\mathbf{x}$.  If $Y \independent Q_{\tau}(Y|\mathbf{X}) | \mathbf{B}_{\tau}^{\top}\mathbf{X}$, where $\mathbf{B}_{\tau}$ is a $p \times d_{\tau}$ matrix, $d_{\tau} \leq p$, then the space spanned by $\mathbf{B}_{\tau}$, denoted by $\mathcal{S}(\mathbf{B}_{\tau})$, is a \textit{$\tau$th quantile dimension reduction subspace} for the regression of $Y$ on $\mathbf{X}$.  This implies that the $d_{\tau} \times 1$ predictor vector $\mathbf{B}_{\tau}^{\top}\mathbf{X}$ contains all the information about $Y$ that is available from $Q_{\tau}(Y|\mathbf{X})$.  The \textit{$\tau$th central quantile subspace} ($\tau$-CQS), denoted by $\mathcal{S}_{Q_{\tau}(Y|\mathbf{X})}$, is defined to be the intersection of all $\tau$th quantile dimension reduction subspaces.  
For the remainder of this paper, we assume that the $\tau$-CQS exists.

The following notation will be used throughout the rest of the paper.  The CS is spanned by the $p \times d$ matrix $\mathbf{A}$, i.e., $\mathcal{S}_{Y|\mathbf{X}}=\mathcal{S}(\mathbf{A})$, and, for a given $\tau$, the $\tau$-CQS is spanned by the $p \times d_{\tau}$ matrix $\mathbf{B}_{\tau}$, i.e., $\mathcal{S}_{Q_{\tau}(Y|\mathbf{X})}=\mathcal{S}(\mathbf{B}_{\tau})$.  The matrices $\mathbf{A}$ and $\mathbf{B}_{\tau}$ are called the \textit{basis} matrices.
It is easy to see that $\mathcal{S}_{Q_{\tau}(Y|\mathbf{X})} \subseteq \mathcal{S}_{Y|\mathbf{X}}$, for any $\tau$.  Therefore, $\mathbf{B}_{\tau}^{\top}\mathbf{X}$ provides a refined structure for the CS, i.e., $\mathbf{B}_{\tau}^{\top}\mathbf{X}= \mathbf{C}_{\tau}^{\top} \mathbf{A}^{\top}\mathbf{X}$, where $\mathbf{C}_{\tau}$ is a $d \times d_{\tau}$ matrix.  

\section{Estimation of the $\tau$-CQS for a SIQR Model} \label{Sec3}

\subsection{Population level}

A SIQR model assumes that $Q_{\tau}(Y|\mathbf{x})=g_{\tau}(\mathbf{B}_{\tau}^{\top}\mathbf{x})$, where $g_{\tau}(\cdot): \mathbb{R} \rightarrow \mathbb{R}$ is an unknown univariate link function, called the nonparametric component, and $\mathbf{B}_{\tau} \in \mathbb{R}^{p}$ is a fixed, but unknown, vector of parameters, called the parametric component.  Since the quantile is assumed to depend on $\mathbf{x}$ only through $\mathbf{B}_{\tau}^{\top}\mathbf{x}$, we sometimes write $Q_{\tau}(Y|\mathbf{B}_{\tau}^{\top}\mathbf{x})$ instead of $Q_{\tau}(Y|\mathbf{x})$.  The SIQR model received particular attention as its nonparametric component is univariate and thus tractable.  For example, Fan et al. (2018) and Christou and Grabchak (2019) used the SIQR model for Value-at-Risk estimation.    

The SIQR model implies that $Y \independent Q_{\tau}(Y|\mathbf{X}) | \mathbf{B}_{\tau}^{\top}\mathbf{X}$ and therefore, assumes a one-dimensional $\tau$-CQS.  The goal is to estimate the parametric component $\mathbf{B}_{\tau}$, which corresponds to the vector of coefficients for the linear combination $\mathbf{B}_{\tau}^{\top}\mathbf{X}$.  Let $R(a_{\tau},\mathbf{b}_{\tau})=E[L\{a_{\tau}+\mathbf{b}^{\top}_{\tau}\mathbf{X},Q_{\tau}(Y|\mathbf{X})\}]$, where $a_{\tau} \in \mathbb{R}$, $\mathbf{b}_{\tau} \in \mathbb{R}^{p}$, and $L\{\theta_{\tau},Q_{\tau}(Y|\mathbf{X})\}$ is a function strictly convex in $\theta_{\tau}$.  Under the SIQR model, and if the conditional expectation $E(\mathbf{b}^{\top}_{\tau}\mathbf{X}| \mathbf{B}_{\tau}^{\top} \mathbf{X})$ is linear in $\mathbf{B}_{\tau}^{\top} \mathbf{X}$ for every $\mathbf{b}_{\tau} \in \mathbb{R}^{p}$, then $\boldsymbol{\beta}_{\tau} \in \mathcal{S}_{Q_{\tau}(Y|\mathbf{X})}$, where 
\begin{eqnarray*} \label{minimum}
(\alpha_{\tau}, \boldsymbol{\beta}_{\tau})= \arg \min_{(a_{\tau}, \mathbf{b}_{\tau})} R(a_{\tau},\mathbf{b}_{\tau}).
\end{eqnarray*}  
This implies that $\boldsymbol{\beta}_{\tau}$ is equal to the coefficients of the linear combination of the predictors up to a multiplier, i.e., $\boldsymbol{\beta}_{\tau}=c\mathbf{B}_{\tau}$ for some $c \in \mathbb{R} \setminus \{0\}$.  This idea comes from the work of Brillinger (1983) and Li and Duan (1989), who considered a similar task but for estimating the one-dimensional CS.       

Moreover, since $\mathcal{S}_{Y|\mathbf{X}}=\mathcal{S}(\mathbf{A})$, then minimizing $R(a_{\tau},\mathbf{b}_{\tau})$ with respect to $a_{\tau}$ and $\mathbf{b}_{\tau}$, is the same as minimizing $R^*(a_{\tau},\mathbf{b}_{\tau})$, where
\begin{eqnarray} \label{new_minimum}
R^*(a_{\tau},\mathbf{b}_{\tau})=E[L\{a_{\tau}+\mathbf{b}^{\top}_{\tau}\mathbf{X},Q_{\tau}(Y|\mathbf{A}^{\top}\mathbf{X})\}].
\end{eqnarray}
Then, $\boldsymbol{\beta}_{\tau}^* \in \mathcal{S}_{Q_{\tau}(Y|\mathbf{X})}$, where  
\begin{eqnarray*} \label{solutions_min}
(\alpha_{\tau}^*, \boldsymbol{\beta}^*_{\tau})= \arg \min_{(a_{\tau}, \mathbf{b}_{\tau})} R^*(a_{\tau},\mathbf{b}_{\tau}).
\end{eqnarray*} 
This allows for an initial dimension reduction using $\mathbf{A}$ for all choices of $\tau$, which is then converted into an estimate of $\mathbf{B}_{\tau}$ for a specific $\tau$.   

\begin{ass} \label{ass}
For a given $\tau$, the conditional expectation $E(\mathbf{b}^{\top}_{\tau}\mathbf{X}| \mathbf{B}_{\tau}^{\top} \mathbf{X})$ is linear in $\mathbf{B}_{\tau}^{\top} \mathbf{X}$ for every $\mathbf{b}_{\tau} \in \mathbb{R}^{p}$.
\end{ass}

\begin{thm} \label{thm3.1}
For a given $\tau \in (0,1)$, assume that $Y \independent Q_{\tau}(Y|\mathbf{X}) | \mathbf{B}_{\tau}^{\top}\mathbf{X}$, where $\mathbf{B}_{\tau}$ is a $p \times 1$ vector.  Under Assumption \ref{ass} and if    
\begin{eqnarray*} \label{solutions_min}
(\alpha_{\tau}^*, \boldsymbol{\beta}^*_{\tau})= \arg \min_{(a_{\tau}, \mathbf{b}_{\tau})} R^*(a_{\tau},\mathbf{b}_{\tau}),
\end{eqnarray*} 
where $R^*(a_{\tau},\mathbf{b}_{\tau})$ is defined in (\ref{new_minimum}), then $\boldsymbol{\beta}_{\tau}^{*} \in \mathcal{S}_{Q_{\tau}(Y|\mathbf{X})}$. 
\end{thm}

\noindent \textbf{Proof:}
Observe that
\begin{eqnarray*}
R^*(a_{\tau},\mathbf{b}_{\tau})&=&E[L\{a_{\tau}+\mathbf{b}^{\top}_{\tau}\mathbf{X},Q_{\tau}(Y|\mathbf{A}^{\top}\mathbf{X})\}]=E[E[L\{a_{\tau}+\mathbf{b}^{\top}_{\tau}\mathbf{X},Q_{\tau}(Y|\mathbf{A}^{\top}\mathbf{X})\}| \mathbf{B}^{\top}_{\tau}\mathbf{X}]]\\
& \geq & E[L[E(a_{\tau}+\mathbf{b}_{\tau}^{\top}\mathbf{X}|\mathbf{B}^{\top}_{\tau}\mathbf{X}),E\{Q_{\tau}(Y|\mathbf{A}^{\top}\mathbf{X})|\mathbf{B}^{\top}_{\tau}\mathbf{X}\}]]\\
&=&E[L\{a_{\tau}+\mathbf{b}^{\top}_{\tau}E(\mathbf{X}|\mathbf{B}_{\tau}^{\top}\mathbf{X}),Q_{\tau}(Y|\mathbf{X})\}]\\
&=&E[L\{a_{\tau}+\mathbf{b}_{\tau}^{\top}P_{\mathbf{B}_{\tau}}(\boldsymbol{\Sigma}_{\mathbf{xx}})^{\top}\mathbf{X},Q_{\tau}(Y|\mathbf{X})\}]\\
&=&E[L\{a_{\tau}+d\mathbf{B}_{\tau}^{\top}\mathbf{X},Q_{\tau}(Y|\mathbf{X})\}]=R^*(a_{\tau},d\mathbf{B}_{\tau}),
\end{eqnarray*}
where $P_{\mathbf{B}_{\tau}}(\boldsymbol{\Sigma}_{\mathbf{xx}})=\mathbf{B}_{\tau}(\mathbf{B}_{\tau}^{\top}\boldsymbol{\Sigma}_{\mathbf{xx}}\mathbf{B}_{\tau})^{-1}\mathbf{B}_{\tau}^{\top}\boldsymbol{\Sigma}_{\mathbf{xx}}$, $\boldsymbol{\Sigma}_{\mathbf{xx}}$ is the covariance matrix of $\mathbf{X}$, and $d$ is a constant. The first inequality follows from Jensen's inequality.  Moreover, the third line follows from the fact that $E\{Q_{\tau}(Y|\mathbf{A}^{\top}\mathbf{X})|\mathbf{B}_{\tau}^{\top}\mathbf{X}\}=E\{Q_{\tau}(Y|\mathbf{X})|\mathbf{B}_{\tau}^{\top}\mathbf{X}\}=E\{Q_{\tau}(Y|\mathbf{B}_{\tau}^{\top}\mathbf{X})|\mathbf{B}_{\tau}^{\top}\mathbf{X}\}=Q_{\tau}(Y|\mathbf{X})$, and the fourth line follows from the fact that under Assumption \ref{ass}, $E(\mathbf{X}|\mathbf{B}_{\tau}^{\top}\mathbf{X})=P_{\mathbf{B}_{\tau}}(\boldsymbol{\Sigma}_{\mathbf{xx}})^{\top}\mathbf{X}$.  
$\blacksquare$

\begin{rmk}
It may be interesting to note that if the distribution of $\mathbf{X}$ is elliptically symmetric, then Assumption \ref{ass} is satisfied for every $\tau \in (0,1)$.  Although the elliptical distribution assumption appears restrictive, among other existing results, Diaconis and Freedman (1984) showed that most low-dimensional projections of high-dimensional data are approximately normal.    
\end{rmk}

An important situation is when the objective function is 
\begin{eqnarray*}\label{OLS}
L\{a_{\tau}+\mathbf{b}^{\top}_{\tau}\mathbf{X},Q_{\tau}(Y|\mathbf{X})\}=\{Q_{\tau}(Y|\mathbf{X})-a_{\tau}-\mathbf{b}^{\top}_{\tau}\mathbf{X}\}^2,
\end{eqnarray*}
which implies the following minimization problem
\begin{eqnarray}\label{OLS_beta}
(\alpha_{\tau}^*, \boldsymbol{\beta}_{\tau}^{*})=  \arg \min_{(a_{\tau},\mathbf{b}_{\tau})} E\{Q_{\tau}(Y|\mathbf{A}^{\top}\mathbf{X})-a_{\tau}-\mathbf{b}^{\top}_{\tau}\mathbf{X}\}^2.
\end{eqnarray}
Theorem \ref{thm3.1} implies that the ordinary least squares (OLS) vector $\boldsymbol{\beta}_{\tau}^{*}$, resulting from regressing $Q_{\tau}(Y|\mathbf{A}^{\top}\mathbf{X})$ on $\mathbf{X}$, belongs to $\mathcal{S}_{Q_{\tau}(Y|\mathbf{X})}$.  
 
\subsection{Sample level - Algorithm 1}

For computational simplicity, we will use the minimization problem (\ref{OLS_beta}) and suggest the following estimation procedure.  First, use a standard dimension reduction technique to estimate $\mathbf{A}$ by $\widehat{\mathbf{A}}$ and form the new $d \times 1$ predictor vector $\widehat{\mathbf{A}}^{\top}\mathbf{X}$.  In this paper, we use the SIR of Li (1991), but one can also use, for instance, the SAVE of Cook and Weisberg (1991) or another technique.  Then, we use data $\{Y_{i}, \mathbf{X}_{i}\}_{i=1}^{n}$ to estimate $\boldsymbol{\beta}_{\tau}^{*}$ by
\begin{eqnarray}\label{OLS_sample}
(\widehat{a}_{\tau},\widehat{\boldsymbol{\beta}}_{\tau})=\arg \min_{(a_{\tau},\mathbf{b}_{\tau})}\sum_{i=1}^{n}\{\widehat{Q}_{\tau}(Y|\widehat{\mathbf{A}}^{\top}\mathbf{X}_{i})-a_{\tau}-\mathbf{b}_{\tau}^{\top}\mathbf{X}_{i}\}^2,
\end{eqnarray}        
where $\widehat{Q}_{\tau}(Y|\widehat{\mathbf{A}}^{\top}\mathbf{X}_{i})$ is a nonparametric estimate of $Q_{\tau}(Y|\widehat{\mathbf{A}}^{\top}\mathbf{X}_{i})$.  There are many ways to estimate $Q_{\tau}(Y|\widehat{\mathbf{A}}^{\top}\mathbf{X}_{i})$; we choose the local linear conditional quantile estimation method introduced in Guerre and Sabbah (2012) as it is simple to implement and tends to work well in practice.  The idea is to take $\widehat{Q}_{\tau}(Y|\widehat{\mathbf{A}}^{\top}\mathbf{X}_{i})=\widehat{q}_{\tau}(\mathbf{X}_{i})$, where
\begin{eqnarray} \label{llqr}
(\widehat{q}_{\tau}(\mathbf{X}_{i}), \widehat{\mathbf{s}}_{\tau}(\mathbf{X}_{i}))=\arg \min_{(q_{\tau}, \mathbf{s}_{\tau})} \sum_{k=1}^{n} \rho_{\tau} \{Y_{k} - q_{\tau} - \mathbf{s}^{\top}_{\tau} \widehat{\mathbf{A}}^{\top}(\mathbf{X}_{k}-\mathbf{X}_{i})\} K \left \{ \frac{\widehat{\mathbf{A}}^{\top}(\mathbf{X}_{k}-\mathbf{X}_{i})}{h} \right\},
\end{eqnarray}   
where $\rho_{\tau}(u)=\{\tau-I(u<0)\}u$.  Here $K(\cdot)$ is a $d$-dimensional kernel function and $h>0$ is a bandwidth.  In this paper, we use a Gaussian kernel and choose the bandwidth using the rule-of-thumb given in Yu and Jones (1998).  Specifically, we select $h=h_{m}[\tau(1-\tau)/\phi\{\Phi^{-1}(\tau)\}^2]^{1/5}$, where $\phi(\cdot)$ and $\Phi(\cdot)$ denote the probability density and cumulative distribution functions of the standard normal distribution, respectively, and $h_{m}$ denotes the optimal bandwidth used in mean regression local estimation.  We now summarize the algorithm.  
  
\hrulefill

\textbf{Sample Level Algorithm 1:} Let $\{Y_{i}, \mathbf{X}_{i}\}_{i=1}^{n}$ independent and identically distributed (iid) observations and fix $\tau \in (0,1)$.
\begin{enumerate}
\item Use the SIR of Li (1991) or a similar dimension reduction technique to estimate the $p \times d$ basis matrix $\mathbf{A}$ of the CS, denoted by $\widehat{\mathbf{A}}$, and form the new sufficient predictors $\widehat{\mathbf{A}}^{\top}\mathbf{X}_{i}$, $i=1,\dots,n$.
\item For each $i=1, \dots, n$, use the local linear conditional quantile estimation method of Guerre and Sabbah (2012) to estimate $Q_{\tau}(Y|\widehat{\mathbf{A}}^{\top}\mathbf{X}_{i})$.  Specifically, take $\widehat{Q}_{\tau}(Y|\widehat{\mathbf{A}}^{\top}\mathbf{X}_{i})=\widehat{q}_{\tau}(\mathbf{X}_{i})$, where $\widehat{q}_{\tau}(\mathbf{X}_{i})$ satisfies (\ref{llqr}).    
\item Take $\widehat{\boldsymbol{\beta}}_{\tau}$ to be
\begin{eqnarray*}
(\widehat{a}_{\tau},\widehat{\boldsymbol{\beta}}_{\tau})=\arg \min_{(a_{\tau},\mathbf{b}_{\tau})}\sum_{i=1}^{n}\{\widehat{Q}_{\tau}(Y|\widehat{\mathbf{A}}^{\top}\mathbf{X}_{i})-a_{\tau}-\mathbf{b}_{\tau}^{\top}\mathbf{X}_{i}\}^2.
\end{eqnarray*}
\end{enumerate}  
Then, $\widehat{\boldsymbol{\beta}}_{\tau}$ defines an estimated basis vector for $\mathcal{S}_{Q_{\tau}(Y|\mathbf{X})}$.  

\hrulefill  

\begin{rmk}
It is sometimes easy to transform $\mathbf{X}$ linearly and study the relation between $Y$ and the transformed predictors.  In general, if $\mathbf{Z}=\mathbf{W}^{\top}\mathbf{X}+\mathbf{b}$ for some invertible matrix $\mathbf{W}$ and some vector $\mathbf{b}$, then $\mathcal{S}_{Q_{\tau}(Y|\mathbf{Z})}=\mathbf{W}^{-1}\mathcal{S}_{Q_{\tau}(Y|\mathbf{X})}$.  This can be proved easily by noting that
\begin{eqnarray*}
Q_{\tau}(Y|\mathbf{X})=Q_{\tau}(Y|\mathbf{B}_{\tau}^{\top}\mathbf{X})=Q_{\tau} \{Y|\mathbf{B}_{\tau}^{\top}\mathbf{W}^{-\top}(\mathbf{Z}-\mathbf{b})\}=Q_{\tau}(Y|\mathbf{B}_{\tau}^{\top}\mathbf{W}^{-\top}\mathbf{Z})=Q_{\tau}\{Y|(\mathbf{W}^{-1}\mathbf{B}_{\tau})^{\top}\mathbf{Z}\}.
\end{eqnarray*}
Therefore, in practice, we can standardize $\mathbf{X}$ to have zero mean and the identity covariance matrix.  We apply the algorithm to $\widehat{\mathbf{Z}}=\widehat{\boldsymbol{\Sigma}}_{\mathbf{xx}}^{-1/2}\{\mathbf{X}-E_{n}(\mathbf{X})\}$, where $\widehat{\Sigma}_{\mathbf{xx}}$ and $E_{n}(\mathbf{X})$ denote the sample covariance matrix and sample mean of $\mathbf{X}$, respectively.  If $\widehat{\boldsymbol{\eta}}_{\tau} \in \mathcal{S}_{Q_{\tau}(Y|\mathbf{Z})}$, where $\mathbf{Z}$ is the population version of $\widehat{\mathbf{Z}}$, then $\widehat{\mathbf{\Sigma}}_{\mathbf{xx}}^{-1/2}\widehat{\boldsymbol{\eta}}_{\tau} \in \mathcal{S}_{Q_{\tau}(Y|\mathbf{X})}$.  
\end{rmk}

\begin{thm}\label{thm5.1} For a given $\tau \in (0,1)$, assume that $Y \independent Q_{\tau}(Y|\mathbf{X}) | \mathbf{B}_{\tau}^{\top}\mathbf{X}$, where $\mathbf{B}_{\tau}$ is a $p \times 1$ vector.  Under Assumption \ref{ass}, Assumptions 1-5 given in Appendix \ref{AppendixA}, and the assumption that $\widehat{\mathbf{A}}$ is $\sqrt{n}$-consistent estimate of the directions of the CS, then $\widehat{\boldsymbol{\beta}}_{\tau}$ is $\sqrt{n}$-consistent estimate of the direction of $\mathcal{S}_{Q_{\tau}(Y|\mathbf{X})}$, where $\widehat{\boldsymbol{\beta}}_{\tau}$ is defined in (\ref{OLS_sample}).    
\end{thm}

\noindent \textbf{Proof:}  See Appendix \ref{AppendixB2}.
$\blacksquare$

\section{Estimation of the $\tau$-CQS for a MIQR Model}\label{Sec4}

\subsection{Population level}

A MIQR model is an extension of a SIQR model, which assumes that $Q_{\tau}(Y|\mathbf{x})=g_{\tau}(\mathbf{B}_{\tau}^{\top}\mathbf{x})$, where $g_{\tau}(\cdot): \mathbb{R}^{d_{\tau}} \rightarrow \mathbb{R}$ is a $d_{\tau}$-dimensional link function, $d_{\tau} \geq 1$, and $\mathbf{B}_{\tau}$ is a $p \times d_{\tau}$ matrix of unknown parameters.  The MIQR model implies that $Y \independent Q_{\tau}(Y|\mathbf{X}) | \mathbf{B}_{\tau}^{\top}\mathbf{X}$, and assumes a $d_{\tau}$-dimensional $\tau$-CQS.  The goal is to estimate the space spanned by the column vectors of $\mathbf{B}_{\tau}$, which correspond to the vectors of coefficients for the linear combinations $\mathbf{B}_{\tau}^{\top}\mathbf{X}$.  If $d_{\tau}$ is strictly greater than 1, then the OLS slope vector $\boldsymbol{\beta}_{\tau}^*$, defined in (\ref{OLS_beta}), is inefficient for estimating $\mathcal{S}_{Q_{\tau}(Y|\mathbf{X})}$, and therefore, a different approach is necessary to produce more vectors in $\mathcal{S}_{Q_{\tau}(Y|\mathbf{X})}$.   

\begin{thm}\label{thm4.1}
For a given $\tau \in (0,1)$, assume that $Y \independent Q_{\tau}(Y|\mathbf{X}) | \mathbf{B}_{\tau}^{\top}\mathbf{X}$, where $\mathbf{B}_{\tau}$ is a $p \times d_{\tau}$ matrix and $d_{\tau} \geq 1$.  Then, under Assumption \ref{ass}, and the assumption that $U_{\tau}$ is a measurable function of $\mathbf{B}_{\tau}^{\top}\mathbf{X}$, $E\{Q_{\tau}(Y|U_{\tau})\mathbf{X}\} \in \mathcal{S}_{Q_{\tau}(Y|\mathbf{X})}$, provided that $Q_{\tau}(Y|U_{\tau})\mathbf{X}$ is integrable.  
\end{thm}

\noindent \textbf{Proof:}
Observe that
\begin{eqnarray*} 
E\{Q_{\tau}(Y|U_{\tau})\mathbf{X}\}&=&E[E\{Q_{\tau}(Y|U_{\tau})\mathbf{X}|\mathbf{B}_{\tau}^{\top}\mathbf{X}\}]=E\{Q_{\tau}(Y|U_{\tau})E(\mathbf{X}|\mathbf{B}_{\tau}^{\top}\mathbf{X})\}\\
&=&E\{Q_{\tau}(Y|U_{\tau})P_{\mathbf{B}_{\tau}}(\boldsymbol{\Sigma}_{\mathbf{xx}})^{\top}\mathbf{X}\}=P_{\mathbf{B}_{\tau}}(\boldsymbol{\Sigma}_{\mathbf{xx}})^{\top}E\{Q_{\tau}(Y|U_{\tau})\mathbf{X}\},
\end{eqnarray*}
where the second line follows from Assumption \ref{ass}.
$\blacksquare$

Suppose that we know one vector $\boldsymbol{\beta}_{\tau,0} \in \mathcal{S}_{Q_{\tau}(Y|\mathbf{X})}$.  The above theorem can be used to find other vectors in $\mathcal{S}_{Q_{\tau}(Y|\mathbf{X})}$ by an \textit{iterative} procedure.  Specifically, for a function $u_{\tau}: \mathbb{R} \rightarrow \mathbb{R}$ and $j=1,\dots,$ $\boldsymbol{\beta}_{\tau,j}=E[Q_{\tau}\{Y|u_{\tau}(\boldsymbol{\beta}_{\tau,j-1}^{\top}\mathbf{X})\}\mathbf{X}] \in \mathcal{S}_{Q_{\tau}(Y|\mathbf{X})}$.  The question now is how to find an initial vector.  Theorem \ref{thm3.1} states that $\boldsymbol{\beta}_{\tau}^{*}$, defined in (\ref{OLS_beta}), belongs to $\mathcal{S}_{Q_{\tau}(Y|\mathbf{X})}$.  Therefore, we set $\boldsymbol{\beta}_{\tau,0}=\boldsymbol{\beta}_{\tau}^{*}$ and, for simplicity, we take $u_{\tau}(t)=t$.    

\begin{cor} \label{cor}
Under the assumptions of Theorem \ref{thm4.1}, the vector $E\{Q_{\tau}(Y|\boldsymbol{\beta}_{\tau}^{*\top}\mathbf{X})\mathbf{X}\}$ belongs to $\mathcal{S}_{Q_{\tau}(Y|\mathbf{X})}$, where $\boldsymbol{\beta}_{\tau}^{*}$ is defined in (\ref{OLS_beta}).
\end{cor}

The above provide a method of forming vectors in the $\tau$-CQS.  Let $\boldsymbol{\beta}_{\tau}^{*}$ as defined in (\ref{OLS_beta}) and set $\boldsymbol{\beta}_{\tau,0}=\boldsymbol{\beta}_{\tau}^{*}$ and, for $j=1,2,\dots, p-1$, $\boldsymbol{\beta}_{\tau,j}=E \{Q_{\tau}(Y|\boldsymbol{\beta}_{\tau,j-1}^{\top}\mathbf{X})\mathbf{X}\}$.  Then, $\boldsymbol{\beta}_{\tau,j} \in \mathcal{S}_{Q_{\tau}(Y|\mathbf{X})}$, $j=0, 1, \dots, p-1$.  However, to obtain linearly independent vectors, we propose the following.  Let $\mathbf{V}_{\tau}$ be the $p \times p$ matrix with column vectors $\boldsymbol{\beta}_{\tau,0}, \dots, \boldsymbol{\beta}_{\tau,p-1}$ and perform an eigenvalue decomposition on $\mathbf{V}_{\tau}\mathbf{V}^{\top}_{\tau}$ to select the $d_{\tau}$ linearly independent eigenvectors $\mathbf{v}_{\tau,1},\dots, \mathbf{v}_{\tau,d_{\tau}}$ corresponding to the $d_{\tau}$ non-zero eigenvalues.  Then, $(\mathbf{v}_{\tau,1},\dots, \mathbf{v}_{\tau,d_{\tau}}) \in \mathcal{S}_{Q_{\tau}(Y|\mathbf{X})}$.  

\subsection{Sample level - Algorithm 2}

 We now summarize the algorithm.  
 
\hrulefill

\textbf{Sample Level Algorithm 2:} Let $\{Y_{i}, \mathbf{X}_{i}\}_{i=1}^{n}$ iid observations and fix $\tau \in (0,1)$.  
\begin{enumerate}
\item Use Algorithm 1 to compute $\widehat{\boldsymbol{\beta}}_{\tau}$, defined in (\ref{OLS_sample}).  Set $\widehat{\boldsymbol{\beta}}_{\tau,0}=\widehat{\boldsymbol{\beta}}_{\tau}$.
\item If $d_{\tau}=1$ stop and report $\widehat{\boldsymbol{\beta}}_{\tau}$ as the estimated basis vector for $\mathcal{S}_{Q_{\tau}(Y|\mathbf{X})}$.  Otherwise, move to Step 3.     
\item Given $j$, where $j=1, \dots, p-1$, 
\begin{enumerate}
\item [(a)] form the predictors $\widehat{\boldsymbol{\beta}}_{\tau,j-1}^{\top}\mathbf{X}_{i}$, $i=1, \dots, n$, and use the local linear conditional quantile estimation method of Guerre and Sabbah (2012) to estimate $Q_{\tau}(Y|\widehat{\boldsymbol{\beta}}_{\tau,j-1}^{\top}\mathbf{X}_{i})$.  Specifically, take $\widehat{Q}_{\tau}(Y|\widehat{\boldsymbol{\beta}}_{\tau,j-1}^{\top}\mathbf{X}_{i})=\widehat{q}_{\tau}(\mathbf{X}_{i})$, where $\widehat{q}_{\tau}(\mathbf{X}_{i})$ is given in (\ref{llqr}), except that we replace $\widehat{\mathbf{A}}$ by $\widehat{\boldsymbol{\beta}}_{\tau,j-1}$.  This leads to a univariate kernel function.   
\item[(b)] let $\widehat{\boldsymbol{\beta}}_{\tau,j}=n^{-1} \sum_{i=1}^{n} \widehat{Q}_{\tau}(Y|\widehat{\boldsymbol{\beta}}_{\tau,j-1}^{\top}\mathbf{X}_{i})\mathbf{X}_{i}$.  
\end{enumerate}
\item Repeat Step 3 for $j=1, \dots, p-1$.   
\item  Let $\widehat{\mathbf{V}}_{\tau}$ be the $p \times p$ matrix with column vectors $\widehat{\boldsymbol{\beta}}_{\tau,j}$, $j=0,1,\dots,p-1$, that is, $\widehat{\mathbf{V}}_{\tau}=(\widehat{\boldsymbol{\beta}}_{\tau,0}, \dots, \widehat{\boldsymbol{\beta}}_{\tau,p-1})$, and choose the eigenvectors $\widehat{\mathbf{v}}_{\tau,k}$, $k=1, \dots, d_{\tau}$, corresponding to the $d_{\tau}$ largest eigenvalues of $\widehat{\mathbf{V}}_{\tau}\widehat{\mathbf{V}}_{\tau}^{\top}$.  Then, 
\begin{eqnarray}\label{CQS_beta}
\widehat{\mathbf{B}}_{\tau}=(\widehat{\mathbf{v}}_{\tau,1}, \dots, \widehat{\mathbf{v}}_{\tau, d_{\tau}})
\end{eqnarray}
is an estimated basis matrix for $\mathcal{S}_{Q_{\tau}(Y|\mathbf{X})}$.
\end{enumerate}

\hrulefill

\begin{rmk}
As in Algorithm 1, for convenience we can work with the standardized predictor $\widehat{\mathbf{Z}}=\widehat{\mathbf{\Sigma}}_{\mathbf{xx}}^{-1/2}\{\mathbf{X}-E_{n}(\mathbf{X})\}$.
\end{rmk}

\begin{thm}\label{thm5.2} For a given $\tau \in (0,1)$, assume that $Y \independent Q_{\tau}(Y|\mathbf{X})|\mathbf{B}_{\tau}^{\top} \mathbf{X}$, where $\mathbf{B}_{\tau}$ is a $p \times d_{\tau}$ matrix and $d_{\tau} \geq 1$.  Under Assumption \ref{ass}, Assumptions 1-5 given in Appendix \ref{AppendixA}, and the assumption that $\widehat{\mathbf{A}}$ is $\sqrt{n}$-consistent estimate of the directions of the CS, then the column vectors of $\widehat{\mathbf{B}}_{\tau}$ are $\sqrt{n}$-consistent estimates of the directions of $\mathcal{S}_{Q_{\tau}(Y|\mathbf{X})}$, where $\widehat{\mathbf{B}}_{\tau}$ is defined in (\ref{CQS_beta})       
\end{thm}

\noindent \textbf{Proof:} See Appendix \ref{AppendixB3}.
$\blacksquare$

\section{Central Subspace for statistical functional $T$}\label{Sec5}

Although the focus of this paper is on the conditional quantile function, the above methodology can be generalized to any statistical functional of interest.  Luo et al. (2014) introduced the $T$-central subspace, denoted by $\mathcal{S}_{T(Y|\mathbf{X})}$, as the smallest subspace spanned by the column vectors of the $p \times d_{T}$ matrix $\mathbf{\Gamma}$ satisfying $Y \independent T(Y|\mathbf{X}) | \mathbf{\Gamma}^{\top} \mathbf{X}$.  Luo et al. (2014) proposed an efficient dimension reduction technique for estimating the fewest linear combinations of $\mathbf{X}$ that contain all the information on the function $T(Y|\mathbf{X})$.  However, the finite sample performance of their proposed efficient estimator is not necessarily the best.  For this reason, we consider an extension of the proposed methodology to any statistical functional.  

\begin{thm} \label{thm_stat}
Assume that $Y \independent T(Y|\mathbf{X}) | \mathbf{\Gamma}^{\top}\mathbf{X}$, where $\mathbf{\Gamma}$ is a $p \times d_{T}$ matrix, $d_{T} \geq 1$.  Under Assumption \ref{ass} and if   
\begin{eqnarray}\label{functional_T}
(\alpha^*, \boldsymbol{\gamma}^{*})=\arg \min_{(a,\boldsymbol{\gamma})} E\{T(Y|\mathbf{\Gamma}^{\top}\mathbf{X})-a-\boldsymbol{\gamma}^{\top}\mathbf{X}\}^2,
\end{eqnarray}
then $\boldsymbol{\gamma}^{*} \in \mathcal{S}_{T(Y|\mathbf{X})}$.  Moreover, if $V$ is a measurable function of $\mathbf{\Gamma}^{\top}\mathbf{X}$, then $E\{T(Y|V)\mathbf{X}\} \in \mathcal{S}_{T(Y|\mathbf{X})}$, provided that $T(Y|V)\mathbf{X}$ is integrable.
\end{thm}

\noindent \textbf{Proof:}  Straightforward extension of the proofs of Theorems \ref{thm3.1} and \ref{thm4.1}.
$\blacksquare$

\begin{cor}
Under the Assumptions of Theorem \ref{thm_stat}, the vector $E\{T(Y|\boldsymbol{\gamma}^{*\top}\mathbf{X})\mathbf{X}\}$ belongs to $\mathcal{S}_{T(Y|\mathbf{X})}$, where $\boldsymbol{\gamma}^{*}$ is defined in (\ref{functional_T}).
\end{cor}

\hrulefill

\textbf{Sample Level Algorithm 3:} Let $\{Y_{i}, \mathbf{X}_{i}\}$ iid observations.
\begin{enumerate}
\item Use SIR of Li (1991) or a similar dimension reduction technique to estimate the $p \times d$ basis matrix $\mathbf{A}$ of the CS, denoted by $\widehat{\mathbf{A}}$, and form the new sufficient predictors $\widehat{\mathbf{A}}^{\top}\mathbf{X}_{i}$, $i=1, \dots, n$.
\item For each $i=1, \dots, n$, estimate $T(Y|\widehat{\mathbf{A}}^{\top}\mathbf{X}_{i})$ using nonparametric techniques.  This step depends on what the function $T$ is.
\item Take $\widehat{\boldsymbol{\gamma}}$ to be
\begin{eqnarray} \label{OLS_T}
(\widehat{a}, \widehat{\boldsymbol{\gamma}})= \arg \min_{(a, \mathbf{c})} \sum_{i=1}^{n} \{ \widehat{T}(Y|\widehat{A}^{\top}\mathbf{X}_{i})-a-\mathbf{c}^{\top}\mathbf{X}_{i}\}^2.
\end{eqnarray} 
\item If $d_{T}=1$ stop and report $\widehat{\boldsymbol{\gamma}}$ as the estimated basis vector for $\mathcal{S}_{T(Y|\mathbf{X})}$.  Otherwise, move to Step 5.
\item Set $\widehat{\boldsymbol{\gamma}}_{0}=\widehat{\boldsymbol{\gamma}}$, where $\widehat{\boldsymbol{\gamma}}$ is defined in (\ref{OLS_T}).
\item Given $j$, for $j=1, \dots, p-1$, 
\begin{enumerate}
\item[(a)] form the predictors $\widehat{\boldsymbol{\gamma}}_{j-1}^{\top}\mathbf{X}_{i}$, $i=1, \dots, n$ and use nonparametric techniques to estimate $T(Y|\widehat{\boldsymbol{\gamma}}_{j-1}^{\top}\mathbf{X}_{i})$.  
\item[(b)] let $\widehat{\boldsymbol{\gamma}}_{j}=n^{-1} \sum_{i=1}^{n} \widehat{T}(Y|\widehat{\boldsymbol{\gamma}}^{\top}_{j-1}\mathbf{X}_{i})\mathbf{X}_{i}$.
\end{enumerate}
\item Repeat Step 6 for $j=1, \dots, p-1$.
\item Let $\widehat{\mathbf{G}}$ be the $p \times p$ matrix with column vectors $\widehat{\boldsymbol{\gamma}}_{j}$, $j=0, 1, \dots, p-1$, that is, $\widehat{\mathbf{G}}=(\widehat{\boldsymbol{\gamma}}_{0}, \dots, \widehat{\boldsymbol{\gamma}}_{p-1})$, and choose the eigenvectors $\widehat{\mathbf{g}}_{k}$, $k=1, \dots, d_{T}$, corresponding to the $d_{T}$ largest eigenvalues of $\widehat{\mathbf{G}}\widehat{\mathbf{G}}^{\top}$.  Then, $\widehat{\boldsymbol{\Gamma}}=(\widehat{\mathbf{g}}_{1}, \dots, \widehat{\mathbf{g}}_{d_{T}})$ is an estimated basis matrix for $\mathcal{S}_{T(Y|\mathbf{X})}$.        
\end{enumerate}

\hrulefill

\begin{rmk} 
Note that the dimension reduction technique used in Step 1 focuses on the entire conditional distribution and performs and initial dimension reduction.  This is then converted into an estimate of $\boldsymbol{\Gamma}$, which now focuses on the statistical functional $T$.  
\end{rmk}      

\section{Structural Dimension} \label{Sec6}

In all the above, we assume that the dimension of a subspace is known.  However, in practice the true dimension of a subspace is unknown and needs to be estimated.  There are several methods proposed for estimating the dimension of a subspace, including a $\mathcal{X}^{2}$-sequential test (Li 1991), a cross validation (CV) criterion (Xia et al. 2002, Wang and Xia 2008), and a Bayesian information criterion (BIC; Zhu et al. 2010).  

The construction of the $\mathcal{X}^{2}$-sequential test can be challenging, white the CV criterion can be computationally expensive.   Therefore, we suggest estimating the dimension of a subspace using the modified BIC-type criterion of Zhu et al. (2010).  The major advantage with this method is that the consistency of the estimator of the relevant matrix is enough to guarantee the consistency of the estimator of the dimension.  

To generalize for any subspace of interest, and without notational confusion, we write $\mathbf{\Lambda}$, with a sample version $\widehat{\mathbf{\Lambda}}$, as a $p \times q$ candidate matrix that targets the subspace of interest.  Let $q$ be the true dimension of the subspace of interest, and $\widehat{q}$ the estimate.  The modified BIC-type criterion is defined as
\begin{eqnarray*}\label{Gnk}
G_{n}(k)=n \frac{\sum_{i=1}^{k} \widehat{\lambda}^{2}_{i}}{\sum_{i=1}^{p} \widehat{\lambda}^{2}_{i}}-C_{n} \left \{ \frac{k(k+1)}{2} \right\},
\end{eqnarray*}
where $\widehat{\lambda}_{1} \geq \dots \geq \widehat{\lambda}_{p}$ are the eigenvalues of the matrix $\widehat{\mathbf{\Lambda}}$, $C_{n}/n \rightarrow 0$ as $n \rightarrow \infty$ and $C_{n} \rightarrow \infty$.  A usual choice for $C_{n}$ is $2n^{3/4}/p$.  Then, $q$ can be estimated by $\widehat{q}= \arg \max_{1 \leq k \leq p} G_{n}(k)$.  In fact, $P(\widehat{q}=q) \rightarrow 1$, under the assumption that $\widehat{\mathbf{\Lambda}}$ is consistent estimate of $\mathbf{\Lambda}$.

\section{Numerical Studies}\label{Sec7}

\subsection{Computational Remarks}

Our proposed methodology consists of two steps.  First, we use Algorithm 1 to obtain the initial value $\widehat{\boldsymbol{\beta}}_{\tau}$, as the OLS slope estimate for the regression of $\widehat{Q}_{\tau}(Y|\widehat{\mathbf{A}}^{\top}\mathbf{X})$ on $\mathbf{X}$.  If the dimension of the $\tau$-CQS is one, we can stop and report $\widehat{\boldsymbol{\beta}}_{\tau}$ as the estimated vector of coefficients for the linear combination $\mathbf{B}_{\tau}^{\top}\mathbf{X}$.  If the dimension of the $\tau$-CQS is greater than one, we set the initial vector $\widehat{\boldsymbol{\beta}}_{\tau,0}$ as $\widehat{\boldsymbol{\beta}}_{\tau}$ and use Algorithm 2 to produce more vectors in $\mathcal{S}_{Q_{\tau}(Y|\mathbf{X})}$.  For the first step, the estimation of the basis matrix $\mathbf{A}$ of the CS is performed using existing consistent dimension reduction techniques.  For the simulations we tried different methods (SIR, SAVE, DR, SIMR), but the results were similar.  Here, we report the results from estimating $\mathbf{A}$ using SIR of Li (1991), where the number of slices is chosen to be max$(10, 2p/n)$.  For the computation of the conditional quantile estimators $\widehat{Q}_{\tau}(Y|\widehat{\mathbf{A}}^{\top}\mathbf{X}_{i})$ and $\widehat{Q}_{\tau}(Y|\widehat{\boldsymbol{\beta}}_{\tau,j-1}^{\top}\mathbf{X}_{i})$, used in Algorithms 1 and 2, respectively, we use local linear conditional quantile estimators, which are computed using the function \texttt{lprq} in the \texttt{R} package \texttt{quantreg}.       

For the estimation accuracy we use the angle between the two subspaces $\widehat{\mathbf{B}}_{\tau}$ and $\mathbf{B}_{\tau}$, where $\widehat{\mathbf{B}}_{\tau}$ denotes an estimate of the $\tau$-CQS with a basis matrix $\mathbf{B}_{\tau}$.  This is used as the measure of the distance between two spaces so that smaller number implies stronger correlation.  The angle is measured in radians, and so we report the value divided by $\pi/2$, and is calculated using the function \texttt{subspace} in the \texttt{R} package \texttt{pracma}.  We call this the estimation error.  We note that we have also tried measuring the distance between two subspaces using the measure proposed by Li et al. (2005), but the results exhibit similar patterns.      

 All simulation results are based on $N=100$ iterations. Unless otherwise stated, the sample size is chosen to be $n=600$, and the quantiles under consideration are $\tau=0.25, 0.5$, and 0.75.

\subsection{Simulation Results} \label{Sec7.2}

\noindent \textbf{Example 1:} We demonstrate the performance of Algorithm 1, where $d_{\tau}=1$, and use $\widehat{\boldsymbol{\beta}}_{\tau}$, defined in (\ref{OLS_sample}), as the estimated basis vector for $\mathcal{S}_{Q_{\tau}(Y|\mathbf{X})}$.           

\noindent \textbf{(a)}  We begin by considering the performance of $\widehat{\boldsymbol{\beta}}_{\tau}$ for different choices of $n$ and $p$.  The data is generated according to the following SIQR model
\begin{eqnarray*}
Y=3X_{1}+X_{2}+\varepsilon,
\end{eqnarray*}
where $\mathbf{X}=(X_{1},\dots,X_{p})^{\top}$ and the error $\varepsilon$ are generated according to a standard normal distribution.  The sample size is given by $n=200, 400$ or 600, and the number of predictors is $p=10, 20$ or 40.  The $\tau$-CQS is spanned by $(3,1,0,\dots,0)^{\top}$, for $\tau=0.25, 0.5$ and 0.75.    The results are given in Table \ref{tab:example1}.  We can observe that the mean estimation error increases with $p$ and decreases with $n$.  Moreover, we observe that the performance of $\widehat{\boldsymbol{\beta}}_{\tau}$ is robust to the specific quantile.  This contradicts the performance of the Luo et al. (2014)'s estimator, where the authors observed that the median CQS performed better than the upper 0.75-CQS (see example (g), Luo et al. 2014).  

\setlength{\arrayrulewidth}{.15em}
\begin{table}[h!]
\caption{\label{tab:example1}\it{Mean (and standard deviation) of the estimation errors for $\widehat{\boldsymbol{\beta}}_{\tau}$, $\tau=0.25, 0.5, 0.75$, for Example 1 (a).}}
\begin{center}
\begin{tabularx}{\textwidth}{cc|@{\extracolsep\fill}ccc}
\hline
$n$ & $p$ & 0.25 & 0.5 & 0.75 \\
\hline
200 & 10 & 0.0529 \ (0.0125) & 0.0530 \ (0.0122) & 0.0534 \ (0.0122)\\
& 20 & 0.0938 \ (0.0169) & 0.0936 \ (0.0165) & 0.0939 \ (0.0167)\\
& 40 & 0.2478 \ (0.0277) & 0.2462 \ (0.0279) & 0.2478 \ (0.0281)\\
400 & 10 & 0.0374 \ (0.0090) & 0.0373 \ (0.0091) & 0.0372 \ (0.0090)\\
& 20 & 0.0537 \ (0.0089) & 0.0538 \ (0.0090) & 0.0539 \ (0.0090)\\
& 40 & 0.0948 \ (0.0114) & 0.0945 \ (0.0114) & 0.0948 \ (0.0116)\\
600 & 10 & 0.0292 \ (0.0080) & 0.0292 \ (0.0080) & 0.0294 \ (0.0081)\\
& 20 & 0.0441 \ (0.0073) & 0.0442 \ (0.0073) & 0.0441 \ (0.0074)\\
& 40 & 0.0693 \ (0.0085) & 0.0690 \ (0.0086) & 0.0691 \ (0.0088)\\
\hline
\end{tabularx}
\end{center}
\end{table}

\noindent \textbf{(b)} We now investigate the performance of $\widehat{\boldsymbol{\beta}}_{\tau}$ for different error distributions.  The data is generated according to the following SIQR models
\begin{eqnarray*}
\text{Model I:} \ Y=X_{1}+X_{2}+X_{3}+X_{4}+\varepsilon,\\
\text{Model II:} \ Y=\exp(X_{1}+X_{2})+\varepsilon,\\
\text{Model III:} \ Y=1+X_{1}+0.4X_{2}+\varepsilon,\\
\text{Model IV:} \ Y=X_{1}/(1+X_{1})^2+\varepsilon,
\end{eqnarray*}
where $\mathbf{X}=(X_{1},\dots,X_{10})^{\top}$ are generated according to a standard normal distribution, and the error $\varepsilon$ is generated according to a standard normal distribution ($\mathcal{N}$), a $t$-distribution with 3 degrees of freedom ($t_{3}$), and a chi-square distribution with 3 degrees of freedom ($\mathcal{X}_{3}^2$).  The $\tau$-CQS is spanned by $(1,1,1,1,0,\dots,0)^{\top}$ for Model I, $(1,1,0,\dots,0)^{\top}$ for Model II, $(1,0.4,0,\dots,0)^{\top}$ for Model III, and $(1,0,\dots,0)^{\top}$ for Model IV, for $\tau=0.25, 0.5$ and 0.75.  Table \ref{tab:example2} demonstrates the mean and standard deviation of the estimation error for the different error distributions and the four models.  We observe that the mean estimation error seems to increase as the symmetry of the error distribution decreases.  Once again, we observe that the performance of the proposed estimator is essentially the same for the different quantiles.  
\setlength{\arrayrulewidth}{.15em}
\begin{table}[h!]
\caption{\label{tab:example2}\it{Mean (and standard deviation) of the estimation errors for $\widehat{\boldsymbol{\beta}}_{\tau}$, $\tau=0.25, 0.5, 0.75$, for Example 1 (b).}}
\begin{center}
\begin{tabularx}{\textwidth}{cc|@{\extracolsep\fill}ccc}
\hline
Model & error & 0.25 & 0.5 & 0.75 \\
\hline
I & $\mathcal{N}$ & 0.0419 \ (0.0107) & 0.0420 \ (0.0106) & 0.0420 \ (0.0106)\\
& $t_{3}$ &  0.0601 \ (0.0142) & 0.0597 \ (0.0140) & 0.0600 \ (0.0141)\\
& $\mathcal{X}_{3}^{2}$ & 0.0815 \ (0.0213) & 0.0823 \ (0.0212) & 0.0830 \ (0.0208)\\
II & $\mathcal{N}$ & 0.1168 \ (0.0283) & 0.1185 \ (0.0272) & 0.1181 \ (0.0268)\\
& $t_{3}$ & 0.1235 \ (0.0279) & 0.1232 \ (0.0274) & 0.1234 \ (0.0276)\\
& $\mathcal{X}_{3}^{2}$ & 0.1405 \ (0.0355) & 0.1398 \ (0.0346) & 0.1394 \ (0.0342)\\
III & $\mathcal{N}$ & 0.0734 \ (0.0177) & 0.0733 \ (0.0178) & 0.0731 \ (0.0178)\\
& $t_{3}$ & 0.1028 \ (0.0278) & 0.1023 \ (0.0279) & 0.1028 \ (0.0280)\\
& $\mathcal{X}_{3}^{2}$ & 0.1305 \ (0.0333) & 0.1315 \ (0.0333) & 0.1334 \ (0.0344)\\
IV & $\mathcal{N}$ & 0.1748 \ (0.0469) & 0.1622 \ (0.0406) & 0.1470 \ (0.0339)\\
& $t_{3}$ & 0.1780 \ (0.0426) & 0.1678 \ (0.0367) & 0.1519 \ (0.0337)\\
& $\mathcal{X}_{3}^{2}$ & 0.1790 \ (0.0454) & 0.1687 \ (0.0399) & 0.1535 \ (0.0360)\\
\hline
\end{tabularx}
\end{center}
\end{table}

\noindent \textbf{(c)}  We investigate the performance of $\widehat{\boldsymbol{\beta}}_{\tau}$ using an $\mathbf{X}$ with dependent components.  The data is generated according to Models I-IV, where $\mathbf{X}=(X_{1},\dots,X_{10})^{\top} \sim \mathcal{N}(\mathbf{0}, (\sigma_{ij})_{1 \leq i,j \leq 10})$ with $\sigma_{ij}=0.5^{|i-j|}$, and the error $\varepsilon$ is generated according to $\mathcal{N}$, $t_{3}$, and $\mathcal{X}_{3}^2$ distributions.  To save space, and since the results follow similar pattern, we only report the results for Model I.  From Table \ref{tab:example3} we observe that the errors are larger than those for $\mathbf{X}$ with independent components.  Further investigation will be considered later - see Example 2 (d).

\setlength{\arrayrulewidth}{.15em}
\begin{table}[h!]
\caption{\label{tab:example3}\it{Mean (and standard deviation) of the estimation errors for $\widehat{\boldsymbol{\beta}}_{\tau}$, $\tau=0.25, 0.5, 0.75$, for Example 1 (c).}}
\begin{center}
\begin{tabularx}{\textwidth}{cc|@{\extracolsep\fill}ccc}
\hline
Model & error & 0.25 & 0.5 & 0.75 \\
 \hline
I & $\mathcal{N}$ & 0.1227 \ (0.0327) & 0.1219 \ (0.0307) & 0.1224 \ (0.0306)\\
& $t_{3}$ & 0.1539 \ (0.0465) & 0.1504 \ (0.0459) & 0.1525 \ (0.0454)\\
& $\mathcal{X}_{3}^{2}$ & 0.2088 \ (0.0567) & 0.2199 \ (0.0576) & 0.2205 \ (0.0602)\\
\hline
\end{tabularx}
\end{center}
\end{table}

\noindent \textbf{(d)} We now evaluate the performance of the modified BIC-type criterion, defined in Section \ref{Sec6}.  The data is generated according to Models I - IV, with the predictors and residual having a standard normal distribution.  We consider the structural dimension of the CS, $d$, to be unknown, and we estimate it using the modified BIC-type criterion.  We apply Algorithm 1 using the $p \times \widehat{d}$ matrix $\widehat{\mathbf{A}}$.  From Table \ref{tab:example4} we observe that the mean estimation errors are very similar to the ones in Table \ref{tab:example2}, suggesting that the dimension of the CS has been consistently estimated using the BIC-type criterion.      

\setlength{\arrayrulewidth}{.15em}
\begin{table}[h!]
\caption{\label{tab:example4}\it{Mean (and standard deviation) of the estimation errors for $\widehat{\boldsymbol{\beta}}_{\tau}$, $\tau=0.25, 0.5, 0.75$, when $d$ is estimated using the BIC-type criterion, for Example 1 (d).}}
\begin{center}
\begin{tabularx}{\textwidth}{c|@{\extracolsep\fill}ccc}
\hline
Model & 0.25 & 0.5 & 0.75 \\
 \hline
I & 0.0414 \ (0.0109) & 0.0413 \ (0.0109) & 0.0413 \ (0.0111)\\
II & 0.1205 \ (0.0404) & 0.1200 \ (0.0396) & 0.1193 \ (0.0388)\\
III & 0.0742 \ (0.0192) & 0.0742 \ (0.0192) & 0.0739 \ (0.0189)\\
IV & 0.1660 \ (0.0495) & 0.1534 \ (0.0363) & 0.1366 \ (0.0298) \\
\hline
\end{tabularx}
\end{center}
\end{table}

\begin{figure}[h!] 
\begin{center}
\includegraphics[width=\textwidth]{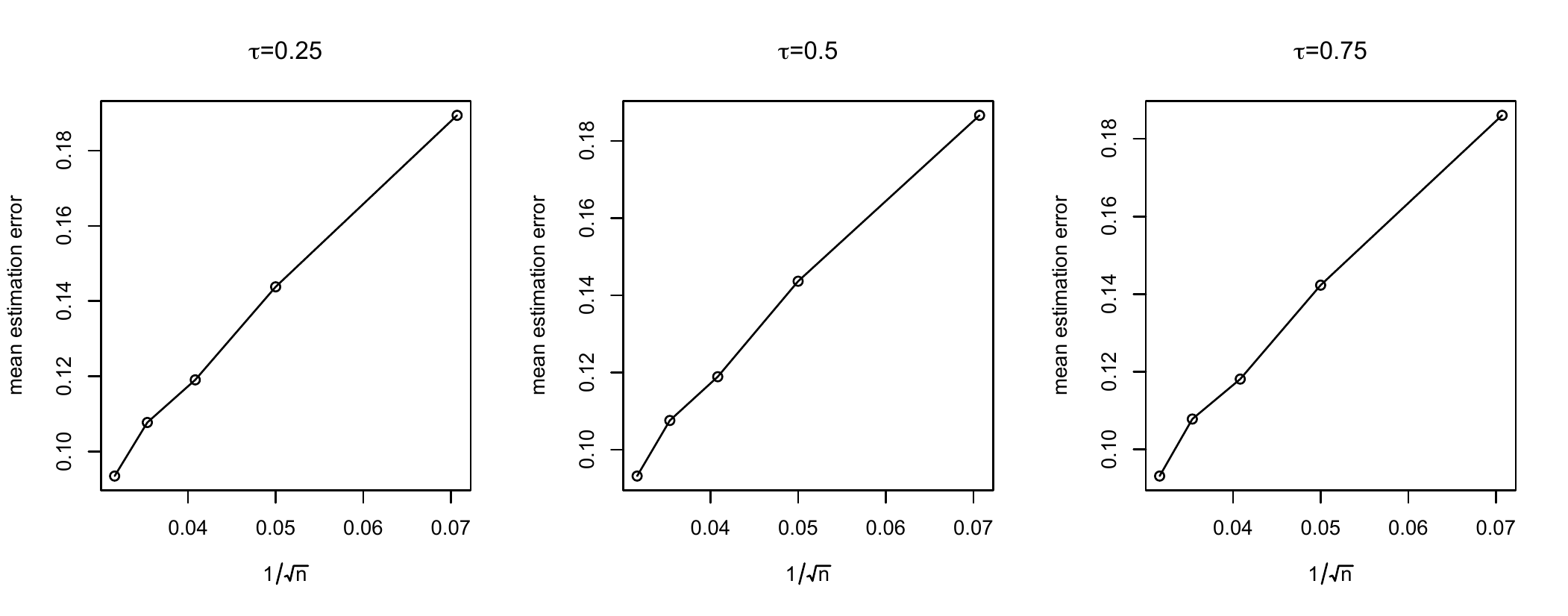}
\caption{\label{fig:example6}{\it{The $\sqrt{n}$-consistency of the proposed estimator $\widehat{\boldsymbol{\beta}}_{\tau}$.}}}
\end{center}
\end{figure}

\noindent \textbf{(e)}  As was already observed in Example 1 (a), the finite sample performance of $\widehat{\boldsymbol{\beta}}_{\tau}$ improves as $n$ increases.  This is due to the $\sqrt{n}$-consistency of the proposed estimator, stated in Theorem \ref{thm5.1}.  We reconsider Model II, where the predictors and the error are generated according to a standard normal distribution.  The sample size is taken to be $n=200, 400, \dots, 1000$.  Figure \ref{fig:example6} shows the observed mean values for the estimation error for the three different quantiles.  The plots clearly indicate an approximate linear relationship between the mean estimation error and $1/\sqrt{n}$, confirming the $\sqrt{n}$-consistency of the proposed estimator $\widehat{\boldsymbol{\beta}}_{\tau}$.  

\noindent \textbf{Example 2:} We demonstrate the performance of Algorithm 2, where $d_{\tau}>1$, and use $\widehat{\mathbf{B}}_{\tau}$, defined in (\ref{CQS_beta}), as the estimated basis matrix for $\mathcal{S}_{Q_{\tau}(Y|\mathbf{X})}$.  We will compare the proposed estimator with the qMAVE procedure of Kong and Xia (2014) and the efficient estimator of Luo et al. (2014).    

\noindent \textbf{(a)}  We begin by considering the performance of $\widehat{\mathbf{B}}_{\tau}$ for different choices of $n$ and $p$.  The data is generated according to the following MIQR model
\begin{eqnarray*}
Y=X_{1}^3+X_{2}+\varepsilon,
\end{eqnarray*}
where $\mathbf{X}=(X_{1},\dots,X_{p})^{\top}$ and the residual $\varepsilon$ are generated according to a standard normal distribution.  The sample size is given by $n=200, 400$ or 600, and the number of predictors is $p=10, 20$ or 40.  The $\tau$-CQS is spanned by $\{(1,0,\dots,0)^{\top}, (0,1,0,\dots,0)^{\top}\}$, for $\tau=0.25, 0.5$ and 0.75.  The results are given in Table \ref{tab:example2a}.  We observe that the mean estimation error increases with $p$ and decreases with $n$.    

\setlength{\arrayrulewidth}{.15em}
\begin{table}[h!]
\caption{\label{tab:example2a}\it{Mean (and standard deviation) of the estimation errors for $\widehat{\mathbf{B}}_{\tau}$, $\tau=0.25, 0.5, 0.75$, using the proposed methodology, for Example 2 (a).}}
\begin{center}
\begin{tabularx}{\textwidth}{cc|@{\extracolsep\fill}ccc}
\hline
$n$ & $p$ & 0.25 & 0.5 & 0.75 \\
\hline
200 & 10 & 0.0552 \ (0.0143) & 0.0552 \ (0.0125) & 0.0562 \ (0.0135) \\
& 20 & 0.0996 \ (0.0187) & 0.0958 \ (0.0192) & 0.0987 \ (0.0202) \\
& 40 & 0.2478 \ (0.0327) & 0.2368 \ (0.0293) & 0.2493 \ (0.0365) \\
400 & 10 & 0.0395 \ (0.0101) & 0.0388 \ (0.0098) & 0.0386 \ (0.0099) \\
& 20 & 0.0604 \ (0.0112) & 0.0594 \ (0.0103) & 0.0616 \ (0.0130) \\
& 40 & 0.1021 \ (0.0146) & 0.0988 \ (0.0143) & 0.1030 \ (0.0187) \\
600 & 10 & 0.0306 \ (0.0086) & 0.0303 \ (0.0085) & 0.0310 \ (0.0088) \\
& 20 & 0.0492 \ (0.0093) & 0.0482 \ (0.0090) & 0.0489 \ (0.0093) \\
& 40 & 0.0786 \ (0.0123) & 0.0744 \ (0.0096) & 0.0764 \ (0.0110) \\
\hline
\end{tabularx}
\end{center}
\end{table}

\noindent \textbf{(b)}  We now compare the performance of the proposed estimator with that of Kong and Xia (2014) and Luo et al. (2014) estimators.  The data is generated according to the following MIQR models 
\begin{eqnarray*}
\text{Model V:} \ Y=X_{1}^3+\exp(X_{2})+\varepsilon\\
\text{Model VI:} \ Y=X_{1}(X_{1}+X_{2}+1)+0.5\varepsilon\\
\text{Model VII:} \ Y=X_{1}/\{0.5+(X_{2}+1.5)^2\}+0.5\varepsilon\\
\text{Model VIII:} \ Y=\cos(3X_{1}/2)+X_{2}^3/2+\varepsilon,
\end{eqnarray*} 
where $\mathbf{X}=(X_{1}, \dots, X_{10})^{\top}$ and the residual $\varepsilon$ are generated according to a standard normal distribution.  The $\tau$-CQS is spanned by $\{(1,0,\dots,0)^{\top}, (0,1,0,\dots,0)^{\top}\}$, for $\tau=0.25, 0.5$ and 0.75, for all models.  Table \ref{tab:example2b} demonstrates the mean and standard deviation of the estimation error of the proposed estimator and of the Kong and Xia (2014) and Luo et al. (2014) estimators.   We observe that the performance of the proposed methodology is comparable with that of the qMAVE procedure of Kong and Xia (2014), while both methods outperform Luo et al. (2014) estimator.  

\setlength{\arrayrulewidth}{.15em}
\begin{table}[h!]
\caption{\label{tab:example2b}\it{Mean (and standard deviation) of the estimation errors for $\widehat{\mathbf{B}}_{\tau}$, $\tau=0.25, 0.5, 0.75$, using the proposed methodology and the Kong and Xia (2014) and Luo et al. (2014) estimators, for Example 2 (b).}}
\begin{center}
\begin{tabularx}{\textwidth}{cc|@{\extracolsep\fill}ccc}
\hline
Model & Methodology & 0.25 & 0.5 & 0.75 \\
\hline
V & Proposed & 0.0672 \ (0.0166) & 0.0644 \ (0.0162) & 0.0657 \ (0.0180)\\
 & Kong and Xia (2014) & 0.0643 \ (0.0170) & 0.0629 \ (0.0150) & 0.0660 \ (0.0165) \\
 & Luo et al. (2014) & 0.1816 \ (0.0473) & 0.2016 \ (0.0617) & 0.2571 \ (0.1003) \\
VI & Proposed & 0.1551 \ (0.0492) & 0.1586 \ (0.0483) & 0.1685 \ (0.0546)\\
 & Kong and Xia (2014) & 0.0978 \ (0.0274) & 0.1001 \ (0.0261) & 0.1053 \ (0.0259) \\
& Luo et al. (2014) & 0.3693 \ (0.1239) & 0.4023 \ (0.1621) & 0.4110 \ (0.1448) \\
VII & Proposed & 0.1108 \ (0.0324) & 0.1091 \ (0.0340) & 0.1125 \ (0.0317)\\
 & Kong and Xia (2014) & 0.1041 \ (0.0258) & 0.1020 \ (0.0231) & 0.1053 \ (0.0245) \\
 & Luo et al. (2014) & 0.6416 \ (0.1420) & 0.6405 \ (0.1361) & 0.6771 \ (0.1401) \\
VIII & Proposed & 0.0894 \ (0.0238) & 0.0874 \ (0.0238) & 0.0899 \ (0.0240) \\
 & Kong and Xia (2014) & 0.1074 \ (0.0301) & 0.1006 \ (0.0233) & 0.1022 \ (0.0229) \\
& Luo et al. (2014) & 0.4539 \ (0.1201) & 0.4591 \ (0.1592) & 0.4843 \ (0.1649) \\
\hline
\end{tabularx}
\end{center}
\end{table}

\noindent \textbf{(c)} We now consider a heteroscedastic model.   The data is generated according to the following MIQR model
\begin{eqnarray*}
Y=X_{1}+X_{2}^3+0.5X_{2}\varepsilon,
\end{eqnarray*}
where $\mathbf{X}=(X_{1}, \dots, X_{10})^{\top}$ and the residual $\varepsilon$ are generated according to a standard normal distribution.  Table \ref{tab:example2heter} demonstrates the mean and standard deviation of the estimation error of the proposed estimator and the Kong and Xia (2014) and Luo et al. (2014) estimators.  As before, the proposed methodology and the qMAVE procedure are comparable, while both methods outperform Luo et al. (2014) estimator.     

\setlength{\arrayrulewidth}{.15em}
\begin{table}[h!]
\caption{\label{tab:example2heter}\it{Mean (and standard deviation) of the estimation errors for $\widehat{\mathbf{B}}_{\tau}$, $\tau=0.25, 0.5, 0.75$, using the proposed methodology and the Kong and Xia (2014) and Luo et al. (2014) estimators, for Example 2 (c).}}
\begin{center}
\begin{tabularx}{\textwidth}{c|@{\extracolsep\fill}ccc}
\hline
 Methodology & 0.25 & 0.5 & 0.75 \\
 \hline
Proposed & 0.0558 \ (0.0177) &0.0543 \ (0.0152) & 0.0576 \ (0.0179) \\
Kong and Xia (2014) & 0.0246 \ (0.0060) & 0.0233 \ (0.0057) & 0.0242 \ (0.0065) \\
Luo et al. (2014) & 0.2935 \ (0.1331) & 0.3208 \ (0.1413) & 0.2512 \ (0.1198)\\
\hline
\end{tabularx}
\end{center}
\end{table} 

\noindent \textbf{(d)}  Here we investigate the performance of the iterative proposed estimator using an $\mathbf{X}$ with dependent components.  The data is generated according to Model V, where $\mathbf{X}=(X_{1},\dots,X_{10})^{\top} \sim \mathcal{N}(\mathbf{0}, (\sigma_{ij})_{1 \leq i,j \leq 10})$ with $\sigma_{ij}=0.5^{|i-j|}$, and the residual $\varepsilon$ is generated according to a standard normal distribution.  From Table \ref{tab:example2c}, we observe that the errors are larger than those for $\mathbf{X}$ with independent components, but the degree by which the mean estimation error of the proposed methodology and of the qMAVE procedure improves upon the Luo et al. (2014)'s method is similar to those for the independent component case. 

\setlength{\arrayrulewidth}{.15em}
\begin{table}[h!]
\caption{\label{tab:example2c}\it{Mean (and standard deviation) of the estimation errors for $\widehat{\mathbf{B}}_{\tau}$, $\tau=0.25, 0.5, 0.75$, using the proposed methodology and the Kong and Xia (2014) and Luo et al. (2014) estimators, for Example 2 (d).}}
\begin{center}
\begin{tabularx}{\textwidth}{cc|@{\extracolsep\fill}ccc}
\hline
Model & Methodology & 0.25 & 0.5 & 0.75 \\
 \hline
V & Proposed & 0.1660 \ (0.0514) & 0.1570 \ (0.0445) & 0.1589 \ (0.0426)\\
 & Kong and Xia (2014) & 0.1326 \ (0.0340) & 0.1292 \ (0.0363) & 0.1328 \ (0.0371) \\
& Luo et al. (2014) & 0.5935 \ (0.1341) & 0.5978 \ (0.1408) & 0.6119 \ (0.1294)\\
\hline
\end{tabularx}
\end{center}
\end{table}

\begin{figure}[h!] 
\begin{center}
\includegraphics[width=\textwidth]{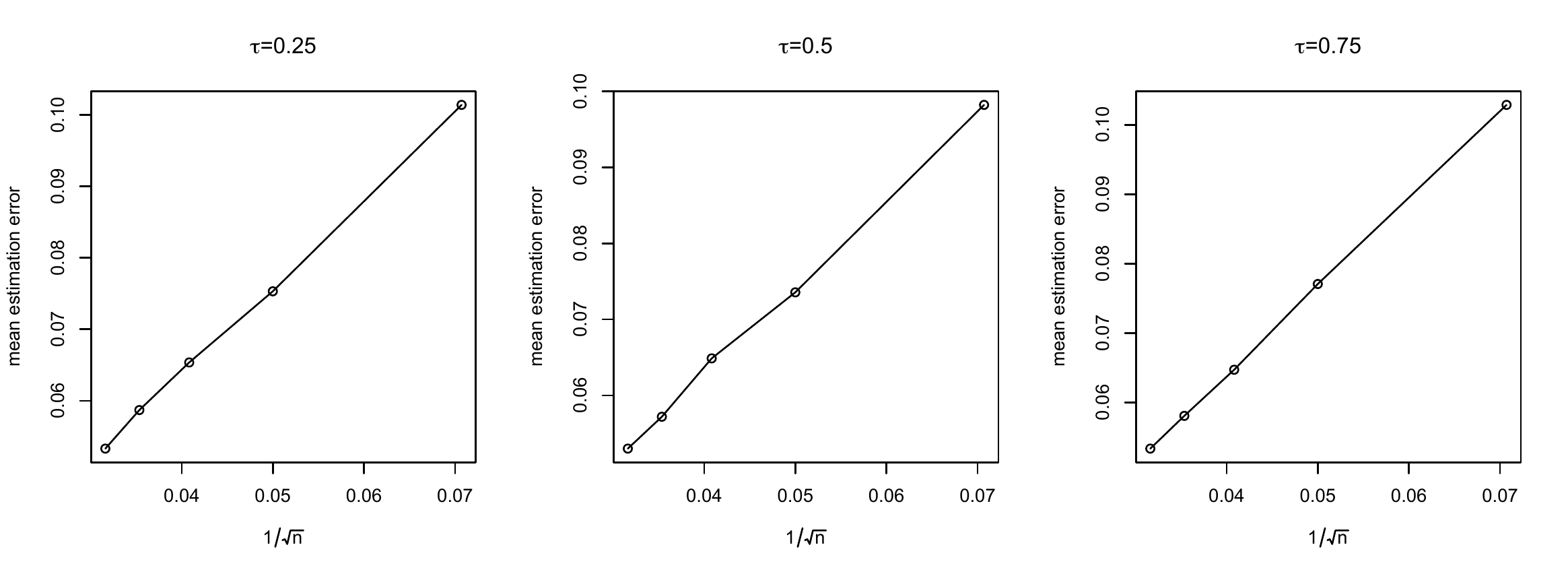}
\caption{\label{fig:example2e}{\it{The $\sqrt{n}$-consistency of the proposed estimator $\widehat{\mathbf{B}}_{\tau}$.}}}
\end{center}
\end{figure}

\noindent \textbf{(e)} Finally, we demonstrate the $\sqrt{n}$-consistency of $\widehat{\mathbf{B}}_{\tau}$, stated in Theorem \ref{thm5.2}.  We reconsider Model V, where the predictors and the error are generated according to a standard normal distribution.  Figure \ref{fig:example2e} shows the observed mean values for the estimation error for the three different quantiles.  As before, the plots clearly confirm the $\sqrt{n}$-consistency of the proposed estimator $\widehat{\mathbf{B}}_{\tau}$.  

\noindent \textbf{Example 3:}  Although the focus of this paper is on the estimation of the $\tau$-CQS, we present one example regarding another statistical functional of interest, the CMS.  Further investigation can be considered in a future paper.  The data is generated according to the model $Y=X_{1}^3+X_{2}\varepsilon$, where the predictors $\mathbf{X}=(X_{1}, \dots, X_{10})^{\top}$ and the residual $\varepsilon$ are generated according to a standard normal distribution.  Observe that the CMS is spanned by $(1,0,\dots,0)^{\top}$, while the CS is spanned by $\{(1,0,\dots,0)^{\top}, (0,1,0,\dots, 0)^{\top}\}$.  The sample size is taken to be $n=600$.  Since the CMS is one-dimensional, we can estimate the coefficients of the linear combination of the predictors using the OLS slope estimate from the regression of $E(Y|\mathbf{A}^{\top}\mathbf{X})$ on $\mathbf{X}$; see (\ref{functional_T}).  The conditional mean $E(Y|\mathbf{A}^{\top}\mathbf{X})$ is estimated nonparametrically using the Nadaraya-Watson estimator with a $d$-dimensional Gaussian kernel function.  Specifically, for the data $\{Y_{i}, \mathbf{X}_{i}\}_{i=1}^{n}$, we use 
\begin{eqnarray*}
\widehat{E}(Y|\widehat{A}^{\top}\mathbf{X}_{i})=\frac{\sum_{k=1}^{n}Y_{k} K\{\frac{\widehat{\mathbf{A}}^{\top}(\mathbf{X}_{k}-\mathbf{X}_{i})}{h}\}}{\sum_{k=1}^{n} K \{\frac{\widehat{\mathbf{A}}^{\top}(\mathbf{X}_{k}-\mathbf{X}_{i})}{h}\}}.  
\end{eqnarray*}
The mean and standard deviation of the estimation error are 0.0233 and 0.0344, respectively.  


\subsection{An application to Upland, CA, Ozone Data} \label{Real_Example}

Air pollution studies are crucial to investigating the effects of various pollutants.  These pollutants can be classified into two categories: the primary pollutants, produced by human activity, and the secondary pollutants, formed under reactions among primary pollutants and other gases.  Since the secondary pollutants are not directly controlled, we would like to investigate their relationship with weather conditions.  Particularly, we consider the relationship between ozone and weather conditions for Upland, CA.  This data set consists of measurements on nine variables.  The dependent variable of interest is the ozone concentration (O$_{3}$) and the other eight variables are: temperature (TMP), inversion base height (InvHt), pressure (PR), visibility (VIS), height (HT), humidity (HUM), inversion base temperature (TMP2), and windspeed (WindSpeed).  The ozone.lsp data set can be downloaded from the Arc package (https://www.stat.umn.edu/arc/software.html).  

Christou (2018) considered this data set and used the SIMR to obtain a robust estimate of the CS.  The author concluded that pressure is the most important variable that affects the ozone concentration, while inversion base height and visibility are also important factors.  Temperature and inversion base temperature also contribute to the variance of ozone concentration.  In this work we apply the proposed methodology to estimate the $\tau$-CQS for different quantile levels.  This investigation is of particular interest in understanding the relationship between extreme values of ozone and weather conditions.  Therefore, we consider estimating the $\tau$-CQS for $\tau=0.1, 0.25, 0.5, 0.75,$ and 0.9.  The scatterplot matrix of the eight predictors does not indicate any serious departures from ellipticity.  Also, the BIC-type criterion suggested that $\widehat{d}_{\tau}=1$, for all $\tau$ under consideration.  The estimated vectors for the $\tau$-CQS are demonstrated in Table \ref{tab:realdata1}. 

From Table \ref{tab:realdata1} we observe that the effects of the weather conditions are essentially stable across the different quantiles.  Pressure is again the most important variable that affects the ozone concentration for all quantile levels, followed by humidity and temperature.  Height and Inversion base height  are also important variables for the $\tau$-CQS, while windspeed has the smallest effect for all quantile levels.   

\setlength{\arrayrulewidth}{.15em}
\begin{table}[h!]
\caption{\label{tab:realdata1}\it{The estimated vectors for the $\tau$-CQS, for $\tau=0.1, 0.25, 0.5, 0.75,$ and 0.9.}}
\begin{center}
\begin{tabularx}{\textwidth}{c|@{\extracolsep\fill}rrrrrrrr}
Direction & TMP & InvHt & PR & VIS & HT & HUM & TMP2 & WindSpeed \\
\hline
$\widehat{\boldsymbol{\beta}}_{0.1}$ & 0.3918 & 0.2271 & -0.6545 & 0.2578 & 0.2213 & 0.4581 & -0.1894 & 0.0730\\
$\widehat{\boldsymbol{\beta}}_{0.25}$ & 0.3921 & 0.2271 & -0.6616 & 0.2578 & 0.2168 & 0.4486 & -0.1936 & 0.0687\\
$\widehat{\boldsymbol{\beta}}_{0.5}$ & 0.3909 & 0.2318 & -0.6683 & 0.2640 & 0.2084 & 0.4383 & -0.1935 & 0.0644\\
$\widehat{\boldsymbol{\beta}}_{0.75}$ & 0.3853 & 0.2352 & -0.6752 & 0.2697 & 0.2043 & 0.4286 & -0.1964 & 0.0587\\
$\widehat{\boldsymbol{\beta}}_{0.9}$ & 0.3776 & 0.2409 & -0.6885 & 0.2798 & 0.1946 & 0.4103 & -0.1950 & 0.0531\\ 
\hline
\end{tabularx}
\end{center}
\end{table}

To compare the proposed methodology with that of Kong and Xia (2014) and Luo et al. (2014), we use the bootstrapped error measurement, introduced in Ye and Weiss (2003).  According to the authors, it is not always straightforward to choose between dimension reduction methods by plotting the response against the corresponding estimated linear combinations.  Instead, they proposed choosing the dimension reduction method that produces an estimated subspace with the smallest variability.  To do that, for each quantile level, we generate 500 bootstrap samples of size 100, and for each sample we compute the proposed estimate $\widetilde{\mathbf{B}}_{\tau}$,  the Kong and Xia (2014) estimate $\widetilde{\mathbf{B}}_{\tau}^{qMAVE}$, and the Luo et al. (2014) estimate $\widetilde{\mathbf{B}}_{\tau}^{LLY}$.  Having also computed the full-sample estimate $\widehat{\mathbf{B}}_{\tau}$, $\widehat{\mathbf{B}}_{\tau}^{qMAVE}$, and $\widehat{\mathbf{B}}_{\tau}^{LLY}$, we compare the three methods by calculating the angle between the bootstrapped subspaces and the full-sample estimate.  
Table \ref{tab:realdata2} reports the mean estimation error for the 500 bootstrap samples for all methods, and for the five different quantile levels.  We observe that the proposed methodology performs much better than that of Kong and Xia (2014) and Luo et al. (2014). 

\setlength{\arrayrulewidth}{.15em}
\begin{table}[h!]
\caption{\label{tab:realdata2}\it{Comparison between proposed methodology, and Kong and Xia (2014) and Luo et al. (2014) estimators, using the bootstrapped error measurement.}}
\begin{center}
\begin{tabularx}{\textwidth}{c|@{\extracolsep\fill}ccccc}
Methodology & 0.1 & 0.25 & 0.5 & 0.75 & 0.9\\
\hline
Proposed & 0.1213 & 0.1155 & 0.1129 & 0.1117 & 0.1114\\
Kong and Xia (2014) & 0.5243 & 0.5137 & 0.5239 & 0.5253 & 0.5270 \\
Luo et al. (2014) & 0.5721 & 0.5649 & 0.5424 & 0.5314 & 0.6066\\
\hline
\end{tabularx}
\end{center}
\end{table}

\section{Discussion} \label{Sec8}

In this work we proposed a new dimension reduction technique with respect to the conditional quantile and suggested an easy to implement algorithm for estimating the $\tau$-CQS, for a given $\tau$.  This method can be further generalized to any statistical functional of interest.  Simulation results and a real data analysis demonstrated the theory developed here and suggested that the proposed methodology has a good finite sample performance, and often outperforms other existing methods.

The presented paper focuses on extracting linear subspaces.  For future work we will consider nonlinear dimension reduction.  Specifically, assuming that $Y \independent Q_{\tau}(Y|\mathbf{X}) | \psi_{\tau}(\mathbf{X})$, where $\psi_{\tau}$ is an arbitrary function, then $\psi_{\tau}(\mathbf{X})$ defines a \textit{nonlinear} sufficient predictor.  The goal is to estimate the nonlinear function $\psi_{\tau}$.

\section{Acknowledgement}
We would like to thank Professors Michael Akritas and Bing Li from the Pennsylvania State University for useful discussions regarding the presented paper. 

\appendix
\section{Notation and Assumptions}\label{AppendixA}

\noindent \textbf{Notation:} We say that a function $m(\cdot): \mathbb{R}^{p} \rightarrow \mathbb{R}$ has the order of smoothness $s$ on the support $\mathcal{X}_{0}$, denoted by $m(\cdot) \in H_{s}(\mathcal{X}_{0})$, if (a) it is differentiable up to order $[s]$, where $[s]$ denotes the lowest integer part of $s$, and (b) there exists a constant $L>0$, such that for all $\mathbf{u}=(u_{1}, \ldots, u_{p})^\top $ with $|\mathbf{u}|=u_{1}+ \cdots +u_{p}=[s]$, all $\tau$ in an interval $[\underline{\tau}, \overline{\tau}]$, and all $\mathbf{x}$, $\mathbf{x}'$ in $\mathcal{X}_{0}$,
\begin{eqnarray*}
|D^{\mathbf{u}}m(\mathbf{x})-D^{\mathbf{u}}m(\mathbf{x'})| \leq L \left\|\mathbf{x}-\mathbf{x}' \right\|^{s-[s]},
\end{eqnarray*}
where $D^{\mathbf{u}}m(\mathbf{x})$ denotes the partial derivative $\partial ^{|\mathbf{u}|}m(\mathbf{x})/\partial x_{1}^{u_{1}} \ldots x_{d}^{u_{d}}$ and $\left\| \cdot \right\|$ denotes the Euclidean norm.

\noindent \textbf{Assumptions}
\begin{enumerate}
\item The following moment conditions are satisfied
\begin{eqnarray*}
E \left\| \mathbf{X}\mathbf{X}^{\top} \right\| < \infty, \ \ E |Q_{\tau}(Y|\mathbf{A}^{\top}\mathbf{X})|^2 < \infty, \ \ E\{Q_{\tau}(Y|\mathbf{A}^{\top}\mathbf{X})^2 \left\| \mathbf{X}\mathbf{X}^{\top} \right\|\}<\infty,
\end{eqnarray*}
for a given $\tau \in (0,1)$.
\item The distribution of $\mathbf{A}^{\top}\mathbf{X}$ has a probability density function $f_{\mathbf{A}}(\cdot)$ with respect to the Lebesgue measure, which is strictly positive and continuously differentiable over the support $\mathcal{X}_{0}$ of $\mathbf{X}$.
\item The cumulative distribution function $F_{Y| \mathbf{A}}(\cdot|\cdot)$ of $Y$ given $\mathbf{A}^{\top}\mathbf{X}$ has a continuous probability density function $f_{Y|\mathbf{A}}(y| \mathbf{A}^{\top}\mathbf{x})$ with respect to the Lebesgue measure, which is strictly positive for $y$ in $\mathbb{R}$ and $\mathbf{A}^{\top}\mathbf{x}$, for $\mathbf{x}$ in $\mathcal{X}_{0}$.  The partial derivative $\partial F_{Y| \mathbf{A}}(y| \mathbf{A}^{\top}\mathbf{x})/ \partial \mathbf{A}^{\top}\mathbf{x}$ is continuous.  There is a $L_{0}>0$, such that
\begin{eqnarray*}
|f_{Y|\mathbf{A}}(y|\mathbf{A}^{\top}\mathbf{x})-f_{Y|\mathbf{A}}(y'|\mathbf{A}^{\top}\mathbf{x}')| \leq L_{0} \left\| (\mathbf{A}^{\top}\mathbf{x},y)-(\mathbf{A}^{\top}\mathbf{x}',y') \right\| \ \text{for all} \ (\mathbf{x},y), (\mathbf{x}',y') \ \text{of} \ \mathcal{X}_{0} \times \mathbb{R}.
\end{eqnarray*}
\item The nonnegative kernel function $K(\cdot)$, used in (\ref{llqr}), is Lipschitz over $\mathbb{R}^{d}$, $d \geq 1$, and satisfies $\int K(\mathbf{z})d \mathbf{z}=1$.  For some $\underline{K}>0$, $K(\mathbf{z}) \geq \underline{K} I\{\mathbf{z} \in B(0,1)\}$ where $B(0,1)$ is the closed unit ball.  The associated bandwidth $h$, used in the estimation procedure, is in $[\underline{h},\overline{h}]$ with $0< \underline{h} \leq \overline{h} < \infty$, $\lim_{n \rightarrow \infty} \overline{h}=0$ and $\lim_{n \rightarrow \infty} (\ln{n})/(n \underline{h}^{d})=0$.
\item $Q_{\tau}(Y|\mathbf{A}^{\top}\mathbf{x})$ is in $H_{s_{\tau}}(\mathcal{T}_{\mathbf{A}})$ for some $s_{\tau}$ with $[s_{\tau}] \leq 1$, where $\mathcal{T}_{\mathbf{A}}=\{\mathbf{z} \in \mathbb{R}^{d}: \mathbf{z}=\mathbf{A}^{\top}\mathbf{x}, \mathbf{x} \in \mathcal{X}_{0}\}$, and $\mathcal{X}_{0}$ is the support of $\mathbf{X}$. 
\end{enumerate}

\section{Proof of Main Results}\label{AppendixB}
\renewcommand{\theequation}{B.\arabic{equation}}
\renewcommand{\thethm}{B.\arabic{thm}}

\subsection{Some Lemmas}\label{Lemmas}

\begin{lemma}\label{LemmaB1}
Under Assumptions 2-5 given in Appendix \ref{AppendixA}, and the assumption that $\widehat{\mathbf{A}}$ is $\sqrt{n}$-consistent estimate of the directions of the CS, then
\begin{eqnarray*}
\sup_{\mathbf{x} \in \mathcal{X}_{0}} |\widehat{Q}_{\tau}(Y|\widehat{\mathbf{A}}^{\top}\mathbf{x})-Q_{\tau}(Y|\mathbf{A}^{\top}\mathbf{x})|=O_{p}(1),
\end{eqnarray*}
where $\widehat{Q}_{\tau}(Y|\widehat{\mathbf{A}}^{\top}\mathbf{x})$ denotes the local linear conditional quantile estimate of $Q_{\tau}(Y|\mathbf{A}^{\top}\mathbf{x})$, given in (\ref{llqr}).  
\end{lemma}

\noindent \textbf{Proof:}
Observe that
\begin{eqnarray*}
\sup_{\mathbf{x} \in \mathcal{X}_{0}} |\widehat{Q}_{\tau}(Y|\widehat{\mathbf{A}}^{\top}\mathbf{x})-Q_{\tau}(Y|\mathbf{A}^{\top}\mathbf{x})| &\leq& \sup_{\mathbf{x} \in \mathcal{X}_{0}} |\widehat{Q}_{\tau}(Y|\widehat{\mathbf{A}}^{\top}\mathbf{x})-\widehat{Q}_{\tau}(Y|\mathbf{A}^{\top}\mathbf{x})|\\
&&+\sup_{\mathbf{x} \in \mathcal{X}_{0}} |\widehat{Q}_{\tau}(Y|\mathbf{A}^{\top}\mathbf{x})-Q_{\tau}(Y|\mathbf{A}^{\top}\mathbf{x})|\\
&=&O_{p}(1).
\end{eqnarray*}
The first term follows from the Bahadur representation of $\widehat{Q}_{\tau}(Y|\widehat{\mathbf{A}}^{\top}\mathbf{x})-\widehat{Q}_{\tau}(Y|\mathbf{A}^{\top}\mathbf{x})$ (see Guerre and Sabbah 2012) and the $\sqrt{n}$-consistency of $\widehat{\mathbf{A}}$.  The second term follows from Corollary 1 (ii) of Guerre and Sabbah (2012).
$\blacksquare$

\textbf{Note:} For the study of the asymptotic properties of $\widehat{\boldsymbol{\beta}}_{\tau}$, defined in (\ref{OLS_sample}), we consider an equivalent objective function.  Observe that minimizing $\sum_{i=1}^{n}\{\widehat{Q}_{\tau}(Y|\widehat{\mathbf{A}}^{\top}\mathbf{X}_{i})-a_{\tau}-\mathbf{b}_{\tau}^{\top}\mathbf{X}_{i}\}^2$ with respect to $(a_{\tau},\mathbf{b}_{\tau})$, is equivalent with minimizing 
\begin{eqnarray}\label{new_objective}
\widehat{S}_{n}(a_{\tau},\mathbf{b}_{\tau})=\frac{1}{2}\sum_{i=1}^{n}\{\widehat{Q}_{\tau}(Y|\widehat{\mathbf{A}}^{\top}\mathbf{X}_{i})-a_{\tau}-\mathbf{b}_{\tau}^{\top}\mathbf{X}_{i}\}^2-\frac{1}{2}\sum_{i=1}^{n}\{\widehat{Q}_{\tau}(Y|\widehat{\mathbf{A}}^{\top}\mathbf{X}_{i})\}^2
\end{eqnarray}
with respect to $(a_{\tau},\mathbf{b}_{\tau})$.  By expanding the square, (\ref{new_objective}) can be written as
\begin{eqnarray} \label{Snhat}
\widehat{S}_{n}(a_{\tau},\mathbf{b}_{\tau})=-(a_{\tau},\mathbf{b}_{\tau})^{\top}\sum_{i=1}^{n}\widehat{Q}_{\tau}(Y|\widehat{\mathbf{A}}^{\top}\mathbf{X}_{i})(1,\mathbf{X}_{i})+\frac{1}{2}(a_{\tau},\mathbf{b}_{\tau})^{\top}\sum_{i=1}^{n}(1,\mathbf{X}_{i})(1,\mathbf{X}_{i})^{\top}(a_{\tau},\mathbf{b}_{\tau}).
\end{eqnarray}  

\begin{lemma}\label{lemma2}
Let $\widehat{S}_{n}(\boldsymbol{\gamma}_{\tau}/\sqrt{n}+(\alpha^*_{\tau},\boldsymbol{\beta}^*_{\tau}))$ be as defined in (\ref{Snhat}), where $\boldsymbol{\gamma}_{\tau}=\sqrt{n} \{(a_{\tau},\mathbf{b}_{\tau})-(\alpha^*_{\tau},\boldsymbol{\beta}^*_{\tau})\}$ and $(\alpha^*_{\tau}, \boldsymbol{\beta}_{\tau}^*)$ is defined in (\ref{OLS_beta}).  Then, under the assumptions of Lemma \ref{LemmaB1} and additionally Assumption 1 of Appendix \ref{AppendixA}, we have the following quadratic approximation, uniformly in $\boldsymbol{\gamma}_{\tau}$ in a compact set,
\begin{eqnarray*}\label{quadr_approx}
\widehat{S}_{n}(\boldsymbol{\gamma}_{\tau}/\sqrt{n}+(\alpha^*_{\tau},\boldsymbol{\beta}^*_{\tau}))= \frac{1}{2}\boldsymbol{\gamma}_{\tau}^{\top}\mathbb{V}\boldsymbol{\gamma}_{\tau}+\mathbf{W}_{\tau,n}^{\top}\boldsymbol{\gamma}_{\tau}+C_{\tau,n}+o_{p}(1),
\end{eqnarray*}
where $\mathbb{V}=E\{(1,\mathbf{X})(1,\mathbf{X})^{\top}\}$, 
\begin{eqnarray} \label{Wn}
\mathbf{W}_{\tau,n}=-\frac{1}{\sqrt{n}}\sum_{i=1}^{n} \widehat{Q}_{\tau}(Y|\widehat{\mathbf{A}}^{\top}\mathbf{X}_{i})(1,\mathbf{X}_{i}),
\end{eqnarray}
and
\begin{eqnarray} \label{Cn}
C_{\tau,n}=-\sum_{i=1}^{n}\widehat{Q}_{\tau}(Y|\widehat{\mathbf{A}}^{\top}\mathbf{X}_{i})(1,\mathbf{X}_{i})^{\top}(\alpha^*_{\tau},\boldsymbol{\beta}^*_{\tau})+\frac{1}{2}(\alpha^*_{\tau},\boldsymbol{\beta}^*_{\tau})^{\top}\sum_{i=1}^{n}(1,\mathbf{X}_{i})(1,\mathbf{X}_{i})^{\top}(\alpha^*_{\tau},\boldsymbol{\beta}^*_{\tau}).
\end{eqnarray}
\end{lemma}

\noindent \textbf{Proof:}  
Observe that
\begin{eqnarray*}
\widehat{S}_{n}(\boldsymbol{\gamma}_{\tau}/\sqrt{n}+(\alpha^*_{\tau},\boldsymbol{\beta}^*_{\tau}))&=& \frac{1}{2n}\boldsymbol{\gamma}_{\tau}^{\top}\sum_{i=1}^{n}(1,\mathbf{X}_{i})(1,\mathbf{X}_{i})^{\top} \boldsymbol{\gamma}_{\tau}-\frac{1}{\sqrt{n}}\sum_{i=1}^{n} \widehat{Q}_{\tau}(Y|\widehat{\mathbf{A}}^{\top}\mathbf{X}_{i})(1,\mathbf{X}_{i})^{\top}\boldsymbol{\gamma}_{\tau}\\
&&-\sum_{i=1}^{n}\widehat{Q}_{\tau}(Y|\widehat{\mathbf{A}}^{\top}\mathbf{X}_{i})(1,\mathbf{X}_{i})^{\top}(\alpha^*_{\tau},\boldsymbol{\beta}^*_{\tau})+\frac{1}{2}(\alpha^*_{\tau},\boldsymbol{\beta}^*_{\tau})^{\top}\sum_{i=1}^{n}(1,\mathbf{X}_{i})(1,\mathbf{X}_{i})^{\top}(\alpha^*_{\tau},\boldsymbol{\beta}^*_{\tau})\\
&=&\frac{1}{2}\boldsymbol{\gamma}^{\top}_{\tau}\mathbb{V}_{n}\boldsymbol{\gamma}_{\tau}+\mathbf{W}_{\tau,n}^{\top}\boldsymbol{\gamma}_{\tau}+C_{\tau,n},
\end{eqnarray*} 
where $\mathbb{V}_{n}=n^{-1}\sum_{i=1}^{n}(1,\mathbf{X}_{i})(1,\mathbf{X}_{i})^{\top}$, and $\mathbf{W}_{\tau,n}$ and $C_{\tau,n}$ are defined in (\ref{Wn}) and (\ref{Cn}), respectively.  It is easy to see that $\mathbb{V}_{n}=\mathbb{V}+o_{p}(1)$, and therefore, 
\begin{eqnarray*}
\widehat{S}_{n}(\boldsymbol{\gamma}_{\tau}/\sqrt{n}+(\alpha^*_{\tau},\boldsymbol{\beta}^*_{\tau}))=\frac{1}{2}\boldsymbol{\gamma}_{\tau}^{\top}\mathbb{V}\boldsymbol{\gamma}_{\tau}+\mathbf{W}_{\tau,n}^{\top}\boldsymbol{\gamma}_{\tau}+C_{\tau,n}+o_{p}(1).
\end{eqnarray*}
Provided that $\mathbf{W}_{\tau,n}$ is stochastically bounded, it follows from the convexity lemma (Pollard 1991) that the quadratic approximation to the convex function $\widehat{S}_{n}(\boldsymbol{\gamma}_{\tau}/\sqrt{n}+(\alpha_{\tau}^*,\boldsymbol{\beta}^*_{\tau}))$ holds uniformly for $\boldsymbol{\gamma}_{\tau}$ in a compact set.  Remains to prove that $\mathbf{W}_{\tau,n}$ is stochastically bounded.

Since $\mathbf{W}_{\tau,n}$ involves the quantity $\widehat{Q}_{\tau}(Y|\widehat{\mathbf{A}}^{\top}\mathbf{X}_{i})$, which is data dependent and not deterministic function, we define 
\begin{eqnarray*}
\mathbf{W}_{\tau,n}(\phi_{\tau})=-\frac{1}{\sqrt{n}}\sum_{i=1}^{n}\phi_{\tau}(Y|\mathbf{A}^{\top}\mathbf{X}_{i})(1,\mathbf{X}_{i}),
\end{eqnarray*}
where $\phi_{\tau}: \mathbb{R}^{d+1} \rightarrow \mathbb{R}$ is a function in the class $\Phi_{\tau}$, whose value at $(y,\mathbf{A}^\top \mathbf{x}) \in \mathbb{R}^{d+1}$ can be written as $\phi_{\tau}(y|\mathbf{A}^{\top}\mathbf{x})$, in the non-separable space $l^{\infty}(y,\mathbf{A}^{\top}\mathbf{x})=\{(y,\mathbf{A}^{\top}\mathbf{x}): \mathbb{R}^{d+1} \rightarrow \mathbb{R}: \left\|\phi _{\tau}\right\|_{(y,\mathbf{A}^{\top}\mathbf{x})}:= \sup_{(y,\mathbf{A}^{\top}\mathbf{x}) \in \mathbb{R}^{d+1}} |\phi_{\tau}(y|\mathbf{A}^{\top}\mathbf{x})|<\infty\}$, and satisfying $E|\phi _{\tau}(Y,\mathbf{A}^{\top}\mathbf{X})|^2 < \infty$ and $E \left\| \phi _{\tau}(Y,\mathbf{A}^{\top}\mathbf{X})^2 \mathbf{X}\mathbf{X}^{\top} \right\| < \infty$.   Since $\Phi_{\tau}$ includes $Q_{\tau}(Y|\mathbf{A}^{\top}\mathbf{x})$, and, according to Lemma \ref{LemmaB1}, includes $\widehat{Q}_{\tau}(Y|\widehat{\mathbf{A}}^{\top}\mathbf{x})$ for $n$ large enough, almost surely, we will prove that $\mathbf{W}_{\tau,n}(\phi_{\tau})$ is stochastically bounded, \textit{uniformly} on $\phi_{\tau} \in \Phi_{\tau}$.

Observe that
\begin{eqnarray*}
\sup _{\phi_{\tau} \in \Phi_{\tau}} \left\| E \left\{\mathbf{W}_{\tau,n}(\phi_{\tau}) \mathbf{W}^{\top}_{\tau,n}(\phi_{\tau}) \right\} \right\| &\leq& \sup _{\phi_{\tau} \in \Phi_{\tau}} \frac{1}{n} \sum_{i=1}^{n} E \left\{ \phi_{\tau}(Y|\mathbf{A}^{\top}\mathbf{X}_{i})^2 \left\| (1,\mathbf{X}_{i})(1,\mathbf{X}_{i})^{\top} \right\| \right\}\\
&=&O[E \left\{ \phi_{\tau}(Y|\mathbf{A}^{\top}\mathbf{X})^2 \left\| (1,\mathbf{X})(1,\mathbf{X})^{\top} \right\| \right\}]=O(1),
\end{eqnarray*}
which follows from the properties of the class $\Phi_{\tau}$ defined above.  Bounded second moment implies that $\mathbf{W}_{\tau, n}(\phi_{\tau})$ is stochastically bounded.  Since
\begin{enumerate}
\item the result was proven uniformly on $\phi_{\tau}$,
\item the class $\Phi_{\tau}$ includes $\widehat{Q}_{\tau}(Y|\widehat{\mathbf{A}}^{\top}\mathbf{x})$ for $n$ large enough, almost surely, and
\item $\mathbf{W}_{\tau,n}(\widehat{Q}_{\tau})=\mathbf{W}_{\tau,n}$, where $\mathbf{W}_{\tau,n}$ is  defined in (\ref{Wn}),
\end{enumerate}
the proof follows.
$\blacksquare$

\subsection{Proof of Theorem \ref{thm5.1}} \label{AppendixB2}

 To prove the $\sqrt{n}$-consistency of $\widehat{\boldsymbol{\beta}}_{\tau}$, enough to show that for any given $\delta_{\tau}>0$, there exists a constant $C_{\tau}$ such that
\begin{eqnarray}\label{positiveprobability}
\Pr  \left \{ \inf_{\left\|\boldsymbol{\gamma}_{\tau} \right\| \geq C_{\tau}}  \widehat{S}_{n}(\boldsymbol{\gamma}_{\tau}/\sqrt{n}+(\alpha^*_{\tau},\boldsymbol{\beta}^*_{\tau}))>\widehat{S}_{n}(\alpha^*_{\tau},\boldsymbol{\beta}^*_{\tau}) \right \} \geq 1-\delta_{\tau},
\end{eqnarray}
where $\widehat{S}_{n}(\boldsymbol{\gamma}_{\tau}/\sqrt{n}+(\alpha_{\tau},\boldsymbol{\beta}_{\tau}))$ defined in (\ref{new_objective}), and implies that with probability at least $1-\delta_{\tau}$ there exists a local minimum in the ball $\{\boldsymbol{\gamma}_{\tau}/\sqrt{n}+(\alpha_{\tau}^*,\boldsymbol{\beta}^*_{\tau}): \left\| \boldsymbol{\gamma}_{\tau} \right\| \leq C_{\tau}\}$.  This in turn implies that there exists a local minimizer such that $\left\|(\widehat{\alpha}_{\tau},\widehat{\boldsymbol{\beta}}_{\tau})-(\alpha_{\tau}^*,\boldsymbol{\beta}^*_{\tau}) \right\|=O_{p}\left(n^{-1/2} \right)$. The quadratic approximation derived in Lemma \ref{lemma2}, yields that
\begin{eqnarray}\label{Difference}
\widehat{S}_{n}(\boldsymbol{\gamma}_{\tau}/\sqrt{n}+(\alpha_{\tau}^*,\boldsymbol{\beta}^*_{\tau}))-\widehat{S}_{n}(\alpha_{\tau}^*,\boldsymbol{\beta}^*_{\tau})=\frac{1}{2}\boldsymbol{\gamma}_{\tau}^\top \mathbb{V}\boldsymbol{\gamma}_{\tau}+\mathbf{W}^\top_{\tau,n}\boldsymbol{\gamma}_{\tau}+o_{p}(1),
\end{eqnarray}
for any $\boldsymbol{\gamma}_{\tau}$ in a compact subset of $\mathbb{R}^{d+1}$.  Therefore, the difference (\ref{Difference}) is dominated by the quadratic term $(1/2)\boldsymbol{\gamma}_{\tau}^\top \mathbb{V}\boldsymbol{\gamma}_{\tau}$ for $\left\|\boldsymbol{\gamma}_{\tau}\right\|$ greater than or equal to sufficiently large $C_{\tau}$.  Hence, (\ref{positiveprobability}) follows. 
$\blacksquare$

\subsection{Proof of Theorem \ref{thm5.2}} \label{AppendixB3}

Let $\widehat{\mathbf{V}}_{\tau}=(\widehat{\boldsymbol{\beta}}_{\tau,0}, \dots, \widehat{\boldsymbol{\beta}}_{\tau,p-1})$ be a $p \times p$ matrix, where $\widehat{\boldsymbol{\beta}}_{\tau,0}=\widehat{\boldsymbol{\beta}}_{\tau}$, defined in (\ref{OLS_sample}), and $\widehat{\boldsymbol{\beta}}_{\tau,j}=E_{n}\{\widehat{Q}_{\tau}(Y|\widehat{\boldsymbol{\beta}}_{\tau,j-1}^{\top}\mathbf{X})\mathbf{X}\}$ for $j=1,\dots,p-1$.  Moreover, let $\mathbf{V}_{\tau}$ be the population level of $\widehat{\mathbf{V}}_{\tau}$.  It is easy to see that $\widehat{\mathbf{V}}_{\tau}$ converges to $\mathbf{V}_{\tau}$ at $\sqrt{n}$-rate.  This follows from the central limi theorem and Lemma \ref{LemmaB1}.  Then, for $\left\| \cdot \right\|$ the Frobenius norm, 
\begin{eqnarray*}
\left\| \widehat{\mathbf{V}}_{\tau} \widehat{\mathbf{V}}_{\tau}^{\top} - \mathbf{V}_{\tau} \mathbf{V}_{\tau}^{\top} \right\| & \leq & \left\| \widehat{\mathbf{V}}_{\tau} \widehat{\mathbf{V}}_{\tau}^{\top} - \widehat{\mathbf{V}}_{\tau} \mathbf{V}_{\tau}^{\top}  \right\| + \left\| \widehat{\mathbf{V}}_{\tau} \mathbf{V}_{\tau}^{\top} - \mathbf{V}_{\tau} \mathbf{V}_{\tau}^{\top} \right\| \\
& =& O_{p}(n^{-1/2}),
\end{eqnarray*}   
and the eigenvectors of $\widehat{\mathbf{V}}_{\tau} \widehat{\mathbf{V}}_{\tau}^{\top}$ converge to the corresponding eigenvectors of $\mathbf{V}_{\tau} \mathbf{V}_{\tau}^{\top}$.  Finally the subspace spanned by the $d_{\tau}$ eigenvectors of $\mathbf{V}_{\tau} \mathbf{V}_{\tau}^{\top}$, falls into $\mathcal{S}_{Q_{\tau}(Y|\mathbf{X})}$ and the proof is complete.
$\blacksquare$

\end{document}